\documentclass[twocolumn,showpacs,pra]{revtex4}
\usepackage{graphicx}
\begin{document}
\draft
\title{Security of the Bennett 1992 quantum-key distribution 
against individual attack over a realistic channel}	
\author{Kiyoshi Tamaki, Masato Koashi, and Nobuyuki Imoto}
\affiliation{CREST Research Team for Interacting Carrier Electronics, \\
 School of Advanced Sciences, The Graduate University for 
Advanced Studies (SOKENDAI),\\
 Hayama, Kanagawa, 240-0193, Japan.
}

\begin{abstract}
The security of two-state quantum key distribution against individual
attack is estimated when the channel has losses and noises. We assume
that Alice and Bob use two nonorthogonal single-photon polarization
states. To make our analysis simple, we propose a modified B92
protocol in which Alice and Bob make use of inconclusive results and
Bob performs a kind of symmetrization of received states. Using this
protocol, Alice and Bob can estimate Eve's information gain as a
function of a few parameters which reflect the imperfections of
devices or Eve's disturbance. 
In some parameter regions, Eve's maximum information gain shows 
counter-intuitive behavior, namely, it decreases as the amount of disturbances increases. For a small noise rate Eve can extract perfect 
information in the case where the angle between Alice's two states is 
small or large, while she cannot extract perfect information for intermediate
 angles.
We also estimate the secret key gain
which is the net growth of the secret key per one pulse. We show the
region where the modified B92 protocol over a realistic channel is
secure against individual attack.
\end{abstract}

\maketitle

\section{Introduction}

Quantum key distribution (QKD) is a way to share between the sender,
Alice, and the receiver, Bob, a secret key whose information 
is not known to the eavesdropper, Eve. Since the first QKD protocol, 
called BB84
protocol, was introduced by Bennett and Brassard \cite{BB84},
 several schemes
for QKD have been proposed, such as Ekert
protocol \cite{ekert}, which is based 
on the nonlocality of quantum mechanics, B92 protocol \cite{B92},
which uses two nonorthogonal states, and so on \cite{imoto}. In each
protocol, since Eve cannot eavesdrop without disturbing quantum
states, she will induce some errors or changes in the transmission rate.
In ideal situations where imperfections such as transmission
losses and dark counting of detectors do not exist, one can prove
that the QKD protocols are secure even when
Eve employs any kind of eavesdropping strategies including so-called
collective attack or coherent attack, which is the most general
attack where a single probe is entangled to the whole transmitted
pulses.
 In practice, however, imperfections exist and if Eve
gets rid of imperfections by using her unlimited technology (for
example, she may replace the noisy channel to a noiseless channel),
 Eve can disturb quantum states to a
certain degree and therefore she obtains some information about
the secret key. Estimation of the amount of information extracted
by Eve and construction of a secure secret key under such situations
are not a trivial problem.

There are several works which deal with the security of realistic
QKD. In \cite{BB841,BB842,BB843,simple,inamori},
it is proven that under some conditions the BB84 protocol is
secure even when Eve employs coherent attack. The security proof
in \cite{simple} reveals a tight connection to
the entanglement purification protocol \cite{EPP} and quantum error
correcting codes, especially Calderbank-Shor-Steane codes
\cite{CSS}. This proof can be also applied to BB84-type protocols such as 
the 6-state protocol \cite{six} where the basis information 
is exchanged over a public channel. 
The security against the individual attack,
which is more restricted but more realistic,
was also discussed for the BB84 protocol \cite{norbert2} 
and for the Ekert protocol \cite{waks}.

On the other hand, we have not yet fully understood 
the security of the B92 protocol. 
The discussions so far concerned with the security against
individual attack \cite{fuchs,slu}. In each
paper, the single-photon
polarization is used as a carrier of the quantum signals, and  
the transmission loss is not included. In \cite{fuchs}, Fuchs and
Peres have calculated the upper bound of Eve's Shannon information
gain averaged over all transmitted bits. Slutsky and co-authers 
have estimated the upper bound of Eve's Renyi information gain in a
B92 protocol as a function of the error rate when Alice and Bob 
discard errors \cite{slu}. The B92 protocol, which uses only two states,
 is conceptually the simplest of all the protocols. In addition, the 
 protocol includes
 a continuous parameter (the nonorthogonality between the two states)
 that can be chosen by the users, forming a striking contrast to
the BB84-type protocols. Further understanding of 
the security of the B92 protocol is thus important not only for 
the practical applications but also for deeper understanding 
of the nature of quantum information.

In this paper, we estimate the security of the B92 protocol
against individual attack over a realistic channel including
noises and transmission losses.
We assume, like in \cite{fuchs,slu}, that Alice encodes the bit values in
 two single-photon polarization states (the original B92 in 
\cite{B92} uses two nonorthogonal coherent states). 
We consider a modified protocol
in which inconclusive measurement results are also used
for the estimation of Eve's information gain,  
in addition to the bit error rate used in the previous analyses.
The problem is also regarded as one of the basic problems of  
quantum eavesdropping, namely, the maximum information leak 
to an adversary who are simulating the noisy 
quantum channel through which Alice transmits Bob a bit 
encoded on two quantum states.

We briefly describe how Alice and Bob can obtain the secret key whose 
information available to Eve is negligibly small. First, Alice prepares 
nonorthogonal single- photon polarization states $|\,0\,\rangle$ and 
$|\,1\,\rangle$, depending on the bit values of the raw key, 
and sends them to Bob.
After propagating through the noisy channel or through eavesdropping
by Eve, $|\,0\,\rangle$ and $|\,1\,\rangle$ generally
become mixed states, $\rho_{0}$ and $\rho_{1}$, respectively.
Secondly, Bob performs measurement whose outcomes
are 0, 1 and ``?''. The outcome ``?'' means an inconclusive result
where Bob cannot determine which state Alice has sent.
Thirdly, Alice and Bob discard inconclusive bits and they
reconcile the errors in conclusive bits by an error correction
protocol which is
the technique of classical information theory to obtain the
reconciled key. Finally, in order to eliminate Eve's information
about the reconciled key, Alice and Bob shorten their keys via a
privacy amplification protocol \cite{privacy amp}, which is also
the technique of the classical information theory. How much they
should shorten the reconciled key depends on how much information
about the reconciled key Eve could have obtained. The last two procedures
are done over a public channel. It is assumed that Eve can listen
to the information exchanged over the public channel, but she
cannot alter it. By means of these techniques, Alice and Bob can
obtain the identical secret key, even under the imperfections and Eve's 
interference. 
 
In the above framework, it is important for Alice and Bob to
estimate Eve's information about the reconciled key properly.
To estimate this information, we propose a slightly modified
protocol. The difference between our protocol and the original
protocol is that when Bob obtains an inconclusive result, 
Alice opens which state she has sent. Using this
extra data, Bob performs data processing which is a kind of
symmetrization of the states received by 
Bob. This makes our analysis simpler without underestimating Eve's
ability.
We can estimate Eve's information by using the symmetrized density
matrices. Eve's information gain is calculated as a function of a
few parameters which characterize the symmetrized density matrices. 
Since Bob's symmetrized density matrices can be 
fully characterized
 by the observed quantities and classical communication, 
the problem we have solved
is equivalent to the problem of how much maximum information Eve can
extract while she simulates the noisy channel. 
The answer to this problem has turned out to be more complex than expected;
We found a counter-intuitive phenomenon where the maximum information
gain is not monotone increasing as a function of the amount of noises
when the angle between Alice's two states is small. We also found that for
a small noise rate, Eve can extract perfect information in the case where
the angle of Alice's two states is very small or very large, while she cannot
extract perfect information for the intermediate angle. We give an 
explanation that is useful to understand these behavior intuitively.

We also performed an optimization of the angle between Alice's two
states that makes the rate of the final key gain as high as possible.
We show that the optimized angle decreases as the amount of noises 
and the transmission losses increase. 
Unfortunately, the assumption 
of the polarization encoding makes the B92 protocol particularly 
weak against channel losses, so that the key gain is always smaller
than in the BB84 protocol. 

This paper is organized as follows. In Sec.~\ref{sec:as}, 
we describe our assumption on the apparatus used by Alice and Bob,
and the limitation on Eve's strategy. Our slightly modified B92
protocol is proposed in Sec.~\ref{secs:protocol}. Then, in 
Sec.~\ref{sec:sym}, we consider Bob's data processing for the
symmetrization of the density matrices, and
in Sec.~\ref{sec:infogain}, we optimize Eve's eavesdropping strategy
and calculate Eve's maximum information gain as a function of
the parameters that reflect imperfections or Eve's disturbance. 
In Sec.~\ref{key-gain}, we calculate the secret key gain as a 
function of these parameters.
Finally Sec.~\ref{summary} is devoted to the summary and discussion.

\section{Assumptions}\label{sec:as} 

In this section, we describe the assumptions on the abilities of the 
legitimate users (Alice and Bob) and the eavesdropper (Eve). We put
these conditions throughout this paper.  
  
First, we describe what Alice and Bob perform 
with ideal apparatus. The B92 protocol is originally considered to
use two nonorthogonal coherent states. In this paper, however, 
we assume two nonorthogonal single-photon polarization states 
for the simplicity of analysis. 
We denote the Hilbert space for each pulse as $H$, which contains
arbitrary photon number states, and the subspace that contains 
only one photon as $H_{1}$. The subspace $H_{1}$ is two-dimensional 
reflecting the polarization degree of freedom, and the states in 
$H_{1}$ are conveniently represented by the Bloch sphere.
We define $|\,\sigma_{\varphi}\,\rangle$ and $|\,\overline
{\sigma_{\varphi}}\,\rangle$ for $-\pi<\varphi<\pi$ as 
follows,
\begin{eqnarray}
|\,\sigma_{\varphi}\,\rangle &\equiv& \cos \frac{\varphi}{2} |\,z + \,\rangle + \sin \frac{\varphi}{2} |\,z - \,\rangle , \nonumber \\
|\,\overline{\sigma_{\varphi}}\,\rangle &\equiv& \sin \frac{\varphi}{2} |\,z + \,\rangle - \cos \frac{\varphi}{2} |\,z - \,\rangle \,\, {\rm for} \, \,\varphi \ge 0
\, , \nonumber \\
{\rm and} \, |\,\overline{\sigma_{\varphi}}\,\rangle &\equiv& - \sin \frac{\varphi}{2} |\,z + \,\rangle + \cos \frac{\varphi}{2} |\,z - \,\rangle \,\, {\rm for} \, \, \varphi < 0 \, ,
\end{eqnarray}
where $|\,z + \,\rangle$ and $|\,z - \,\rangle$ are the eigenstates of $\sigma_{z}$ ($z$ component of Pauli matrix) whose eigenvalues are $+ 1$ and $- 1$, respectively. $\varphi$ is the angle between $|\sigma_{\varphi} \rangle$ and $|\,z+\,\rangle$ on the $x-z$ plane in the Bloch sphere. 

\begin{figure}
 \begin{center}
 \includegraphics[scale=0.6]{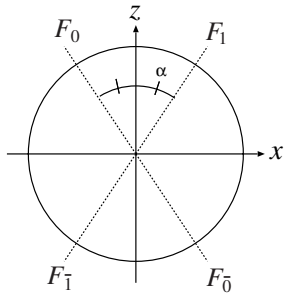}
 \caption{Bob's measurement basis in the Bloch sphere. \label{fig1} }
 \end{center}
\end{figure}

We assume that Alice prepares the following states depending 
on the bit value,                        
\begin{equation}
| \, 0 \rangle \equiv | \, \sigma_{- \alpha'} \rangle \,\, {\rm and}\,| \, 1 \rangle \equiv | \,  \sigma_{\alpha'} \,\rangle  \quad (0 \leq \alpha' \leq \pi/2),
 \label{Alice's choice} 
\end{equation}
where the parameter $\alpha'$ characterizes the nonorthogonality 
between the two states, such that
\begin{equation}
\langle0|1\rangle = \cos \alpha' \,.
\label{Alice's choice2}
\end{equation}
When these states are sent to Bob through the noisy quantum channel,
$|\,0\,\rangle$ and $|\,1\,\rangle$ generally
 become mixed states, $\rho_{0}$ and $\rho_{1}$, respectively. 
On these states, 
Bob measures the polarization on the basis 
$\{| \, \sigma_{- \alpha} \rangle,| \, \overline{\sigma_{- \alpha}} \rangle\}$ or $\{| \, \sigma_{\alpha} \rangle,| \, \overline{\sigma_{\alpha}} \rangle\}$, which is selected randomly. The whole measurement process is 
described by the following POVM \cite{sPOVM}
\begin{eqnarray}
 F_{0} &\equiv& \frac{1}{2}\,| \, \sigma_{- \alpha} \rangle \langle \sigma_{- \alpha} | \nonumber \\
 F_{\overline{0}} &\equiv& \frac{1}{2}\,| \, \overline{\sigma_{- \alpha}} \rangle \langle \overline{\sigma_{- \alpha }} | \nonumber \\
 F_{1} &\equiv& \frac{1}{2}\,| \, \sigma_{\alpha} \rangle \langle \sigma_{\alpha} | \nonumber \\
 F_{\overline{1}} &\equiv& \frac{1}{2}\,| \, \overline{\sigma_{\alpha}} \rangle \langle \overline{\sigma_{\alpha}} | \nonumber \\
 F_{{\rm V}} &\equiv& 1 - F_{0} - F_{\overline{0}} - F_{1} - F_{\overline{1}}
 \label{povm}
\end{eqnarray}
where ``V'' means the  states which contain zero or more than one photon,
where allowances are made for the effect of transmission losses.
$F_{{i}}$ and $F_{\overline{i}}$ ($i = 0,1$) are shown 
in Fig.~\ref{fig1} schematically using Bloch sphere.
Here we allow general cases where $\alpha$ is not necessarily equal to 
$\alpha'$.   We call the events inconclusive where Bob's outcome of the measurement is $0$ or $1$, and the events where Bob's outcome is $\overline{0}$ or 
$\overline{1}$ conclusive. 

\begin{figure}
 \begin{center}
 \includegraphics[scale=0.7]{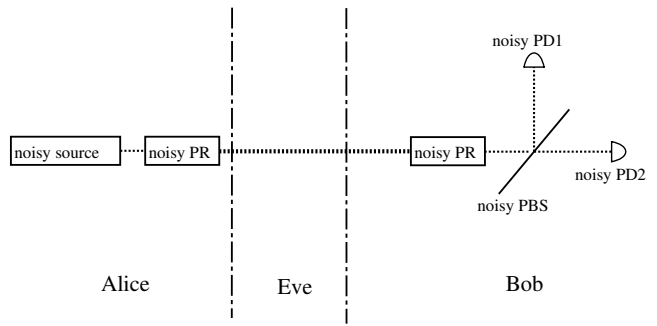}
 \caption{An example of experimental setup. ``noisy PR'' is the noisy polarization rotator, ``noisy PBS'' is a noisy polarization beam splitter, and ``noisy PD1'' and ``noisy PD2'' are noisy photon detectors. Our assumption is that this setup can be transformed into Fig.~\ref{fig3}.\label{fig2}}
 \end{center}
\end{figure}

\begin{figure}
 \begin{center}
 \includegraphics[scale=0.6]{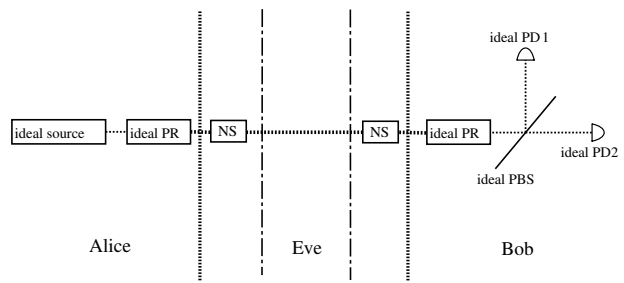}
 \caption{ The model of our setup. The ``NS'' represents the noise source of Alice or Bob's device. Since Alice and Bob cannot separate noises due to imperfections and ones due to Eve, they should assume, for security, that Eve can also control NS, which extends Eve's region up to dotted line. Due to this assumption, the legitimate users' apparatuses look like ideal. ``Ideal PR'' is an ideal polarization rotator which accurately rotates the polarization of photons, ``ideal PBS'' is an ideal polarization beam splitter, and ``ideal PD 1 (2)'' is an ideal photon detector with unit quantum efficiency and no dark counting. \label{fig3}}
 \end{center}
\end{figure}

In reality, the transmission line and Alice and Bob's apparatus is not perfect
(see Fig.~\ref{fig2}). We assume 
one condition on the character of the imperfection, namely,
the imperfection is equivalently represented as a noise source 
placed just after Alice's ideal apparatus, and a noise source 
just before Bob's ideal apparatus. Under this condition, 
Fig.~\ref{fig2} becomes Fig.~\ref{fig3}. 
We need some care for this assumption. For example, if the noisy 
photon detectors PD1 and PD2 in Fig.~\ref{fig2} have different quantum 
efficiencies, we cannot transform this model into Fig.~\ref{fig3} directly. 
However,
if Bob interchanges  noisy PD 1 and noisy PD 2 at random, which effectively makes the efficiency of these detectors identical,
we can transform Fig.~\ref{fig2} into Fig.~\ref{fig3}, and the 
assumption can be met.

The benefit of the above assumption is that we can use a simpler
model in which Alice and Bob's apparatus 
is perfect, in the following sense. In Fig.~\ref{fig3},
the region bounded by dash-dotted lines is under Eve's control.
Suppose that this region is extended up to the dotted lines.
While this assumption may make the length of the final key
shorter than the optimum one,
at least we can avoid the risk of underestimating Eve's ability.
In this model, every imperfection is attributed to the property 
of the quantum channel, and the analysis is considerably simplified.
This assumption has been also used in the previous works 
\cite{waks,esti}.

 We assume that Eve's eavesdropping is restricted to
  individual attack.
The definition of individual attack is that eavesdropping is 
independently done for each pulse and there exists no 
correlation of events among pulses. In the most general individual
attack,
Eve prepares her auxiliary system (probe) E in an initial state 
$|w \rangle_{\rm E}$, 
interacts it with one pulse sent by Alice via a unitary operation $U$, and 
performs measurements on the probe E to obtain the bit
 information. We do not impose any restriction on Eve's 
 measurement on the probe. We do not treat more general attack,
  such as collective attack or coherent attack.

\section{Protocol}\label{secs:protocol} 

In this section, we propose a protocol slightly modified from the
original one, which makes it possible to identify the density matrices
of the states received by Bob more tightly.
The main idea is that Alice and Bob can identify the density matrices 
more precisely by monitoring more parameters.
In our protocol, Alice and Bob monitor not only the error rate in conclusive
bits, but also the statistics of inconclusive bits. This additional information makes it possible to identify the density matrices after a symmetrization which is described in Sec.~\ref{sec:sym}.

Our modified protocol consists of the following steps. 

\begin{enumerate}
 \item Alice determines a bit value $j=0$ or $j=1$ randomly, and sends to Bob the single-photon polarization state
$|\, j \rangle$ defined by Eq.~(\ref{Alice's choice}).
 
 \item Bob performs the measurement described by POVM $\{F_\mu\}(\mu=0,1,\bar{0},\bar{1},{\rm V})$
defined by Eq.~(\ref{povm}). 
  \item If the measurement result was  $\mu=\bar{1}$, Bob sets the
  received bit value as $j^\prime=0$, and if
$\mu=\bar{0}$, he sets $j^\prime=1$. In both cases, Bob tells Alice over the public channel that the result was
conclusive, and Alice adopts $j$ and Bob adopts $j^\prime$ as a bit value of their raw keys, respectively.
If the measurement result was  $\mu=0,1,$ or ${\rm V}$, Bob opens the result $\mu$, and Alice opens the bit value $j$
in the cases $\mu=0,1$.
 \item Alice and Bob repeat steps 1--3 $n_{\rm total}$ times and obtain their raw keys.

 \item{Error reconciliation:} To make an identical key (the reconciled key) from their raw keys, Alice and Bob
       perform an error correction protocol.
 The reconciled key consists of ``correct bits'' which were found to be correct and
 ``flipped bits'' which were found to be incorrect and hence flipped in the 
error correction protocol.
 In order not to leak the additional information about the final key to Eve, 
a previously shared secret key is used to encrypt the communication 
over the public channel. A small length of the secret key is 
also used to make sure that the reconciled key is identical,
using an authentication protocol \cite{authen}.
 
 \item{Estimation of Eve's information:}
After the public communication about the inconclusive events in step 3 and 
the error reconciliation in step 5, Bob knows Alice's original bit values
 for the cases where
Bob's  measurement results were $\mu=0,1,\bar{0}, \bar{1}$.
He can determine the number of events $n_{j,\mu}$ $(\mu=0,1,\bar{0}, \bar{1};
j=0,1)$ where Alice's choice was 
$j$ and Bob's result was $\mu$.
 Assuming that $n_{\rm total}$ is large, $2 n_{j,\mu}/n_{\rm total}$
gives a good estimate of the quantity 
 ${\rm Tr}(F_\mu\rho_j)$. Bob estimates the maximum information
that can be leaked to Eve under the condition
\begin{equation}
{\rm Tr}(F_\mu\rho_j)=2 n_{j,\mu}/n_{\rm total}\,.
\label{constraint}
\end{equation}
 The estimation of Eve's information gain is separately made for
 the correct bits and for the flipped bits.
 \item{Privacy Amplification:} To eliminate the information leaked to Eve, Alice and Bob produce a secure final key by shortening the correct bits and 
 flipped bits of reconciled key according to the information gains 
 estimated in step 6 via a privacy amplification protocol. 

\end{enumerate}

The most important part in the quantum key distribution through a noisy 
channel is the estimation of 
the leaked information done in step 6. The detailed procedures 
in the proposed protocol above were chosen so as to make the estimation 
easier. First, we use not only the error rate in conclusive bits 
but also the measurement results of inconclusive bits to fix 
the density operator of the state delivered to Bob more tightly.
Secondly, the optimization of Eve's measurement on her 
probe system is simplified by estimating Eve's information about
 correct bits and flipped bits independently. For the correct bits,
 the state of Eve's probe when the bit value of the final key is $0$ is
 a pure state  
$C_{0}\langle \overline{1}\,|\,U
  \,|\,0\,\rangle \,|\,w\,\rangle_{\rm E} \equiv | \, \phi_{0} \rangle_{\rm E}$, and
 the state when the bit value is $1$ is $C_{1}\langle \overline{0}\,|U
  \,|\,1\,\rangle \,|\,w\,\rangle_{\rm E} \equiv | \, \phi_{1} \rangle_{\rm E}$, where
$C_{0}$ and $C_{1}$ are the constants for normalization.

In our estimation of Eve's information gain on each correct bit,
we use the information gain $\rm{I}$ defined through 
the collision probability $P(i|\nu)^2$ averaged over possible
Eve's outcomes $\nu$. Here $P(i|\nu)$ is the posteriori probability that
Alice's bit value is $i$ (the state of Eve's probe was 
$| \, \phi_{i} \rangle_{\rm E}$) 
under the condition that Eve has obtained outcome
$\nu$.
The explicit definition of the information gain $\rm{I}$ is
\begin{eqnarray}
\rm{I} \equiv 1 -  \left[-\log_{2} \sum_{i=0,1} \sum_{\nu}
P(\nu) P(i|\nu)^2\right]  \, ,
\end{eqnarray}
where $P(\nu)$ is the probability that Eve obtains 
outcome $\nu$. 
This measure is useful since it directly tells us 
how much we should shorten the key in the privacy
amplification \cite{esti,privacy amp}.
The quantity $\rm{I}$ is maximized by 
the Von Neumann 
measurement on the basis symmetrically arranged around
the vectors $ |\, \phi_{0} \rangle_{\rm E}$ and $| \, \phi_{1} \rangle_{\rm E}$
\cite{slu,opt-measure}, and 
 thus we obtain the maximum information gain for each correct
bit ${\rm I}^{{\rm Gc}}$ which can be written as 
\begin{eqnarray}
{\rm I}^{{\rm Gc}}=\log_{2}\left(2-|Q|^2\right)  \, ,
 \label{IR} 
\end{eqnarray}
where 
\begin{equation}
 Q \equiv  {}_{\rm E} \langle\,\phi_{0}\,|\,\phi_{1}\,\rangle_{\rm E} = 
\frac{{}_{\rm E}\langle w|\langle 0 | U^{\dagger}| \overline{1} \rangle\langle
\overline{0} | U | 1 \rangle \,|w \rangle_{\rm E} }{\sqrt{|{}_{\rm E}\langle w|
\langle 0|U^{\dagger}|\overline{1}\rangle|^{2} } \sqrt{|\langle \overline{0} |U|1\rangle \,|w\rangle_{\rm E} |^{2}}} \, . \label{QQ}
\end{equation}
Note that the same measurement also maximizes the information 
gain measured with respect to Shannon entropy \cite{opt-measure},
and its maximum value  
${\rm I}^{{\rm Gc}}_{{\rm S}}$
is given by
\begin{equation}
 {\rm I}^{{\rm Gc}}_{{\rm S}}= 1- h\left(\frac{1 - \sqrt{1 - |Q|^{2}}}{2}\right) \,,
 \label{IS} 
\end{equation}
where $h(x)$ is the entropy function defined as
$h(x) \equiv -x\log_{2}x-(1-x)\log_{2}(1-x)$.

The maximum information gain ${\rm I}^{{\rm Gf}}$ for each
flipped bit can be obtained by
merely changing the definition of $Q$ with $Q' \equiv {}_{\rm E}\langle\, \phi_{0}
'\,|\,\phi_{1} '\,\rangle_{\rm E}$, where $| \, \phi_{0}' \rangle_{\rm E} \equiv
C_{0}'  \langle
\overline{0}\,|\,{\rm U}
  \,|\,0\,\rangle \,|\,w\,\rangle_{\rm E} \equiv$ and
   $ | \, \phi_{1}' \rangle_{\rm E} \equiv C_{1}' \langle \overline{1}\,|\,{\rm
U}
  \,|\,1\,\rangle \,|\,w\,\rangle_{\rm E}$.
The problem of finding the maxima of $\rm{I}^{{\rm Gc}}$ and 
$\rm{I}^{{\rm Gf}}$ reduces to finding the minimum of $|Q|$ or $|Q'|$, which will be solved in
Sec.~\ref{sec:infogain}.

 In our analysis, we assume, for simplicity, an unrealistic assumption
 that Eve can deal correct bits and flipped bits independently,
 which may overestimate the amount of the leaked information. 
 If we try to deal these together, we need to optimize Eve's measurement
 in four dimensional Hilbert space, 
and the optimization would be more complicated.
We leave the tight estimation of the leaked information in such cases 
to future investigations.

We have to be careful about leaked information during the error reconciliation protocol. The redundant information used in step 5 contains the partial information of the secret key. So, in order to prevent this leakage of information, redundancy bits are encrypted by previously shared secret key.
On the other hand, since we assume Eve employs
 individual attack, in other words the events for different pulses
  are independent,
 the public communication concerned with inconclusive results
 in step 4 does not leak any information about the conclusive results, and
 so we need not to encrypt this communication.

\section{Symmetrization of Eve's strategy}\label{sec:sym}

In this section, we consider the symmetrization of Eve's strategy
which simplifies our analysis, but never underestimates Eve's ability.
The discussion can be similarly applied to the flipped bits.

The eavesdropping strategy of Eve, specified by the unitary 
operator $U$, gives the information gain about the correct bits
of the final key that is determined through the quantity $Q(U)$
defined by Eq.~(\ref{QQ}). It also determines the density 
operators $\rho_j$ ($j=0,1$) of the states delivered to Bob as follows,

\begin{equation}
\rho_j = {\rm Tr}_{\rm E} \left[ U|j\rangle|w\rangle_{\rm E} {}_{\rm E}\langle 
w|\langle j|U^{\dagger} \right] \, .
\label{density}
\end{equation}

On the other hand, these density matrices must satisfy 
Eq.~(\ref{constraint}). The problem to be solved is to find the
minimum value of $|Q|$ under the constraint Eq.~(\ref{constraint}).

A standard way of symmetrization would be to replace the transformation
$U$ to an operation with higher symmetry $\tilde{U}$, such that 
Eqs.~(\ref{constraint}) and (\ref{density}) still hold, and $|Q(U)|\ge 
|Q(\tilde{U})|$ holds.
Then, Eve's information gain  can be estimated by minimization of 
$|Q|$ over possible $\tilde{U}$ without fear of underestimation.

\begin{figure}
 \begin{center}
 \includegraphics[scale=0.6]{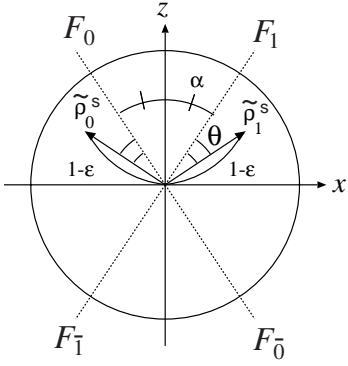}
 \end{center}
\caption{The density operators on $x$-$z$ plane in the Bloch sphere that Bob uses to estimate Eve's extracted information. $\tilde \rho_{0}^{\rm s}$ and $\tilde \rho_{1}^{\rm s}$ do not have $y$ components and they are symmetric with respect to $z$ axis. \label{fig4}}
\end{figure}

Instead of using such a standard scheme, here we invoke a transformation
$U\rightarrow U^{\rm s}$ with $|Q(U)|\ge |Q(U^{\rm s})|$, but
Eqs.~(\ref{constraint}) and (\ref{density}) with $U$ replaced by 
$U^{\rm s}$ are not 
necessarily satisfied. The transformation $U\rightarrow U^{\rm s}$
is explained in Appendix A.  The states that would be 
received by Bob when Eve performed $U^{\rm s}$ are given by  
\begin{equation}
\rho_{j}^{\rm s}={\rm Tr}_{\rm E} \left[ U^{\rm s}|j\rangle|w
\rangle_{\rm E}{}_{\rm E}\langle w|\langle j|(U^{\rm s})^{\dagger} \right].
\label{defrhosj}
\end{equation}
Due to the symmetry of $U^{\rm s}$, these states are simply parametrized by
$\{\theta,\epsilon, T\}$ as follows:
\begin{eqnarray}
 \rho_{0}^{\rm s} &=& T \left(1 - \frac{\epsilon}{2} \right) \left| \, \sigma_{- (\alpha +
\theta)} \rangle
\langle \sigma_{- (\alpha + \theta)} \right| \nonumber \\ &+&  T \, \frac{\epsilon}{2} | \, \overline{\sigma_{-
(\alpha + \theta)}} \rangle \langle \overline{\sigma_{- (\alpha + \theta)}}  | + ( 1 - T )
| {\rm vac} \rangle \langle {\rm vac} | \, ,\nonumber\\ \label{den1}
\end{eqnarray}
and
\begin{eqnarray} 
 \rho_{1}^{ \rm s} &=& T \left(1 - \frac{\epsilon}{2} \right) \left| \, \sigma_{(\alpha +
\theta)} \rangle
\langle \sigma_{(\alpha + \theta)} \right| \nonumber \\ &+&  T \, \frac{\epsilon}{2} | \,
\overline{\sigma_{(\alpha + \theta)}} \rangle \langle \overline{\sigma_{(\alpha + \theta)}}  | + ( 1
- T ) |{\rm vac} \rangle \langle {\rm vac}  | \, \label{den2}\,,
\end{eqnarray} 
where $|{\rm vac}\rangle$ is the vacuum state. These density matrices
are shown in Fig.~\ref{fig4}.
An important point is that the states $\rho_{j}^{\rm s}$ are
related to $\rho_{j}$ as in Eqs.~(\ref{A1})--(\ref{A5}),
and hence, the parameters $\{\theta,\epsilon, T\}$ 
should be related to the actually observed quantities
as follows:
\begin{eqnarray}
(1-\epsilon)\cos\theta={\rm Tr}[(F_0-F_{\bar{0}})\rho_{0}^{\rm s}]
\nonumber \\
=\frac{1}{2}({\rm Tr}[(F_0-F_{\bar{0}})\rho_{0}]+
{\rm Tr}[(F_1-F_{\bar{1}})\rho_{1}])
\nonumber \\
=(n_{0,0}-n_{0,\bar{0}}+n_{1,1}-n_{1,\bar{1}})/n_{\rm total} 
\label{eqepth}
\end{eqnarray}
\begin{eqnarray}
(1-\epsilon)\cos(\theta+2\alpha)={\rm Tr}[(F_1-F_{\bar{1}})\rho_{0}^{\rm s}]
\nonumber \\
=\frac{1}{2}({\rm Tr}[(F_1-F_{\bar{1}})\rho_{0}]+
{\rm Tr}[(F_0-F_{\bar{0}})\rho_{1}])
\nonumber \\
=(n_{0,1}-n_{0,\bar{1}}+n_{1,0}-n_{1,\bar{0}})/n_{\rm total}
\end{eqnarray}
and
\begin{eqnarray}
T= 1-{\rm Tr}[F_{\rm V}(\rho_0+\rho_1)/2]
\nonumber \\
=\sum_{j=0,1}
\sum_{\mu=0,1,\bar{0},\bar{1}}n_{j,\mu}/n_{\rm total}.
\label{eqT}
\end{eqnarray}
Eqs. (\ref{den1})-(\ref{eqT}) imply that the density operators
$\rho_{0}^{\rm s}$ and $\rho_{1}^{\rm s}$ can be completely 
specified by Bob through the observed quantities. 

The above argument is summarized as follows. Suppose that Eve 
conducted a strategy $U$ and Bob obtained the values of  
$\{\theta,\epsilon,T\}$ through the observed quatities and 
relations (\ref{eqepth})-(\ref{eqT}).
The states $\rho_{j}^{\rm s}$ is determined by 
Eqs. (\ref{den1}) and (\ref{den2}).
Then, there exists an attack with unitary operator $U^{\rm s}$
satisfying Eq.(\ref{defrhosj}) and $|Q(U)|\ge |Q(U^{\rm s})|$.
Hence, the minimum of $|Q(U)|$ can be safely estimated 
by the minimization of $|Q(U^{\rm s})|$ under the condition 
that $U^{\rm s}$ satisfies Eq.(\ref{defrhosj}).

Let us further reduce the minimization problem in a 
convenient form.
Using Schmidt decomposition, Eqs.~(\ref{den1}) and (\ref{den2})
lead to the expressions of the total pure states  

\begin{eqnarray}
U^{\rm s} | \, 0 \rangle \,| \, w \rangle_{\rm E} &=& \sqrt{T \, \left( 1 -
\frac{\epsilon}{2}\right)} \,  | \, \sigma_{-(\alpha + \theta)} \,\rangle \,| \, a_{1} \,\rangle_{\rm E}  \nonumber \\ &+& \sqrt{T \, \frac{\epsilon}{2}} | \, \overline{\sigma_{-
(\alpha + \theta)}} \rangle \,| \, a_{2} \,\rangle_{\rm E} \nonumber \\
 &+& \sqrt{1 - T} | \, {\rm vac} \rangle \,| \, a_{\rm v} \rangle_{\rm E} 
\, , 
\label{Schmidt1}
\end{eqnarray}
and
\begin{eqnarray}
U^{\rm s} | \, 1 \rangle \,| \, w \rangle_{\rm E} &=& \sqrt{T \, \left( 1 -
\frac{\epsilon}{2}\right)} \,  | \, \sigma_{(\alpha + \theta)} \,\rangle \,| \, b_{1} \,
\rangle_{\rm E}  \nonumber \\ &+& \sqrt{T \, \frac{\epsilon}{2}} | \,
\overline{\sigma_{(\alpha + \theta)}} \rangle \,| \, b_{2} \,\rangle_{\rm E}
\nonumber \\
 &+& \sqrt{1 - T} | \, {\rm vac} \rangle \,| \, b_{\rm v} \rangle_{\rm E}
 \, ,  
 \label{Schmidt2}
\end{eqnarray}
 where $| \, a_{i} \rangle_{\rm E}$ and $| \, b_{i} \rangle_{\rm E} (i=1,2)$ 
are orthogonal to the space spanned by $| \, a_{\rm v} \rangle_{\rm E}$ and 
$| \, b_{\rm v} \rangle_{\rm E}$, since Eve knows the photon number in this 
strategy.
$| \, a_{i} \rangle_{\rm E}$ and $| \, b_{i} \rangle_{\rm E} (i=1,2)$ satisfy   
\begin{equation}
 _{\rm E}\langle a_{i} | \, a_{j} \rangle_{\rm E} = \delta_{i,j}, \quad _{\rm E}\langle b_{i} | \, b_{j} \,
\rangle_{\rm E} = \delta_{i,j} \, ,
\end{equation}
and hence connected by a unitary operator $\hat{\xi}$, namely,
\begin{equation}
 | \, b_{i} \rangle_{\rm E}=\hat{\xi}| \, a_{i} \rangle_{\rm E}.
\end{equation}
By applying the discussion in Appendix \ref{apA}, the unitary 
operator $\hat{\xi}$ is a real orthogonal matrix in the basis 
${\rm W}$, since $U^{\rm s}$, $|\sigma_{\varphi}\rangle$, 
$|\overline{\sigma_{\varphi}}\rangle$,
 $|{\rm vac}\rangle$, and $|w\rangle$ are invariant under $Z_{\rm W}$. 

The condition for $U^{\rm s}$ to be unitary can be written as
\begin{equation}
 _{\rm E}\langle w | \langle 0 | ( U^{\rm s})^{\dagger} U^{\rm s} | \, 1 \,
\rangle \,| \, w \rangle_{\rm E} = \langle 0|1 \rangle \, ,
\end{equation}
which means
\begin{equation}
T{\rm Tr}\left[ \hat B \hat \xi \right]+ (1 - T) \cos\phi = \cos \alpha' ,
\label{innertemp}
\end{equation}
where
\begin{eqnarray}
\cos\phi&=&_{\rm E}\langle a_{\rm v} | \, b_{\rm v} \rangle_{\rm E}, \\
\hat B &=& \sum_{i, j} B_{i j} |\,a_{i}\,\rangle_{\rm E} {}_{\rm E}\langle
\,a_{j}|, \\
B_{11} &=& \left(1 - \frac{\epsilon}{2}\right)\,\cos\left( \alpha +\theta \right) \, , \\
B_{12} &=& B_{21} = \sqrt{\left(1 - \frac{\epsilon}{2}\right)\,\frac{\epsilon}{2}}\,\sin\left(\alpha +\theta \right) \, , \\
\end{eqnarray}
and
\begin{eqnarray}
B_{22} = - \, \frac{\epsilon}{2}\,\cos\left(\alpha +\theta\right). 
\end{eqnarray}  
Note that for arbitrary $\phi$ and arbitrary real orthogonal matrix 
$\hat{\xi}$ satisfying Eq.~(\ref{innertemp}), there exists a corresponding 
unitary operator $U^{\rm s}$.

Using $\hat{\xi}$ and Eq.~(\ref{QQ}), $Q(U^{\rm s})$ can be written as a function of $\hat{\xi}$ as
\begin{equation}
Q(\hat{\xi}) = {\rm Tr}
\left[ \hat A \hat \xi \right]
\label{Qdayo}
\end{equation}
where
\begin{eqnarray}
\hat A &=& \sum_{i, j} A_{i j} |\,a_{i}\,\rangle_{\rm E} {}_{\rm E}\langle
\,a_{j}|, \\
A_{11} &=& \frac{\left(2 - \epsilon\right)\, \sin^{2} \left(\alpha + \frac{\theta}{2}\right)}{\left[1- (1-\epsilon) \, \cos (2 \alpha + \theta) \right] }\, ,\label{hatA}\\
A_{12} &=& A_{21} = - \frac{\sqrt{\epsilon \left(2 - \epsilon\right)}\,\sin \left(2 \alpha + \theta \right)}{2\left[1- (1-\epsilon) \, \cos (2 \alpha + \theta) \right]}\, , 
\end{eqnarray}
and
\begin{eqnarray}
A_{22} = \frac{\epsilon \, \cos^{2} \left(\alpha + \frac{\theta}{2}\right)}{\left[1-(1-\epsilon) \, \cos (2 \alpha + \theta) \right]} . \label{Qc} 
\end{eqnarray}  
Therefore, the problem of finding the minimum of $|Q(U)|$ reduces to 
the problem of finding the minimum of $|Q(\hat{\xi})|$ under the 
constraint
\begin{equation}
 \left| T{\rm Tr}
\left[ \hat B \hat \xi \right] - \cos \alpha' \right|\le 1-T.
\label{rinner}
\end{equation}

\section{Optimization of Eve's apparatus and her information gain}
\label{sec:infogain}

 In this section, we determine the minimum of $|Q(\hat{\xi})|$ under the 
constraint Eq.~(\ref{rinner}). It is convenient to introduce
a function $Q_{\rm min}(B)$ which is defined as the minimum of
$|Q(\hat{\xi})|$ under the condition
\begin{equation}
B(\hat{\xi})\equiv 
{\rm Tr}
\left[ \hat B \hat \xi \right]=B
\label{Bda}
\end{equation}
for $-B_{\rm max}\le B \le B_{\rm max}$, where $B_{\rm max}$ is the 
maximum of ${\rm Tr}[ \hat B \hat \xi ]$ over all $\hat{\xi}$ for a
given $\hat B$. 
Since $B_{11}B_{22}-B_{12}^{2}\le0$, $B_{\rm max}$ is 
given by 
\begin{equation}
B_{\rm max}^{2}=(B_{11}-B_{22})^2+4B_{12}^2\,.
\end{equation}
Note that the relations $B(\hat{-\xi})=-B(\hat{\xi})$ and 
$|Q(-\hat{\xi})|=|Q(\hat{\xi})|$ imply that
$Q_{\rm min}(B)=Q_{\rm min}(-B)$.
Several properties of $Q_{\rm min}(B)$ are derived from the 
fact that $Q_{\rm min}(\cos\alpha^\prime)$ 
is equal to the minimum of $|Q(\hat{\xi})|$ under the 
constraint Eq.~(\ref{rinner}) when  $T=1$. 
Since Eve can freely determine the bit value $j$
when $\cos\alpha^\prime=0$, we have $Q_{\rm min}(0)=0$. 
Moreover, since 
Eve can always convert the initial states $|j\rangle$ to 
$|j^\prime\rangle$ such that 
$\langle 1^\prime | 0^\prime \rangle\ge \cos\alpha'$,
$Q_{\rm min}(B)$ is a never-decreasing continuous function for
$B\ge0$.
Therefore, $Q_{\rm min}(B)=0$ in a range $|B| \le B_0$, and 
$Q_{\rm min}(B)>0$ for $B_0<|B| \le B_{\rm max}$.
The minimum of $|Q(\hat{\xi})|$ under the 
constraint Eq.~(\ref{rinner}) is given by 
\begin{equation}
\min |Q|= 0 \;\;\; {\rm when}\,\,\frac{\cos\alpha^\prime-
(1-T)}{T}\le B_0  
\end{equation}
and
\begin{eqnarray}
&&\min |Q|= Q_{\rm min}\left(\frac{
\cos\alpha^\prime-(1-T)}{T}\right) \nonumber \\
&&{\rm when}\,\,\frac{
\cos\alpha^\prime-(1-T)}{T}> B_0 \,.
\end{eqnarray}
As we discuss the detail in Appendix B, in order to minimize $|Q|$,
operator $\hat{\xi}$ must satisfy 
\begin{equation}
[(\nu\hat A - \lambda \hat B)^2,\hat{\xi}]=0.
\label{koyuu}
\end{equation}
where $\nu$ and $\lambda$ are real parameters which satisfy 
[$(\nu,\lambda)\neq (0,0)$].

The form of operator $\hat{\xi}$ satisfying 
this condition depends on the rank of 
$\nu\hat A - \lambda \hat B$.
For the moment, we assume that $\epsilon\neq 0$. Since 
the rank of $\hat A$ is $1$ and  that of $\hat B$
is $2$, the rank of 
$\nu\hat A - \lambda \hat B$ is $1$ or $2$.

\begin{figure}
 \begin{center}
 \includegraphics[scale=0.9]{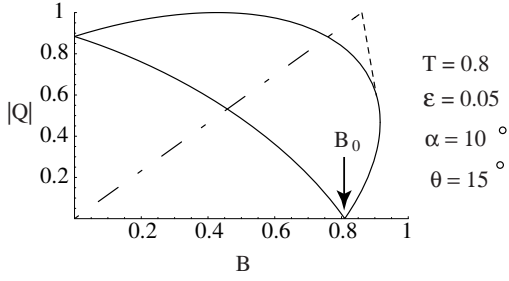}
 \caption{$|Q|$ vs $B$ when $\alpha= 10^{\circ}$, $\epsilon = 0.05$, $\theta = 15^{\circ}$, and $T=0.8$. The solid line represents $|Q|$ obtained by Type 1, the dot-dashed line represents Type 2, and the dotted line represents Type 3. In this figure, $|Q_{\rm min}| = 0$ for $0 \le B \le B_0$ and $|Q_{\rm min}| =|Q_{\rm min}^{(1-)}|$ for $B_0 \le B \le B_{\rm max}$. \label{BQ}}
 \end{center}
\end{figure}

First, we consider the case where the rank of $\nu\hat A - \lambda \hat B$
is $2$.
Let $H_0$ be the subspace 
spanned by $|a_1\rangle_{\rm E}$ and $|a_2\rangle_{\rm E}$, which is  
 the support of $\nu\hat A - \lambda \hat B$, and $H_0^{\perp}$ be
the complementary space of $H_0$. Let 
$\hat P_{H_0}$ and $\hat P_{H_0^{\perp}}$ be the projection operators to 
the corresponding subspaces.
$\hat P_{H_0} [(\nu\hat A - \lambda \hat B)^2,\hat{\xi}] 
\hat P_{H_0}^{\perp}=0$ 
implies that $\hat \xi|a_1\rangle_{\rm E}\in H_0$ and 
$\hat \xi|a_2\rangle_{\rm E}\in H_0$. 
Hence the matrix form of $\hat \xi$ in $H_0$ 
with respect to the basis $\{|a_1\rangle_{\rm E}, |a_2\rangle_{\rm E}\}$
should be written as
\begin{equation}
\hat \xi=\pmatrix{\cos\eta &\sin\eta \cr \sin\eta & -\cos\eta},
\label{type1}
\end{equation}
which we call Type 1, or
\begin{equation}
\hat \xi=\pmatrix{\cos\eta & -\sin\eta \cr \sin\eta & \cos\eta},
\label{type2}
\end{equation}
which we call Type 2.
With the help of Eq.(\ref{Qdayo}), $Q(\hat{\xi})$ and $B(\hat{\xi})$
for Type 1 are 
given by functions of parameter $\eta$ as 
\begin{equation}
Q^{(1)}(\eta)=(A_{11}-A_{22})\cos\eta+2 A_{12} \sin\eta
\end{equation}
and from Eq.(\ref{Bda})
\begin{equation}
B^{(1)}(\eta)=(B_{11}-B_{22})\cos\eta+2 B_{12} \sin\eta,
\label{B1}
\end{equation}
respectively.
Similarly, for Type 2, we have 
\begin{equation}
Q^{(2)}(\eta)=(A_{11}+A_{22})\cos\eta=\cos\eta
\end{equation}
and
\begin{eqnarray}
B^{(2)}(\eta)&=&(B_{11}+B_{22})\cos\eta
\nonumber \\
&=&(1-\epsilon)\cos(\alpha +\theta)\cos\eta.
\label{B2}
\end{eqnarray}
$|Q^{(1)}|$ and $|Q^{(2)}|$ are plotted in Fig.~\ref{BQ} by solid line and
dot-dashed line respectively. 
Note that the choice of $\hat\xi$
\begin{eqnarray}
 \hat\xi|\,a_{1}\,\rangle_{\rm E} = \cos \eta |\,a_{1}\,\rangle_{\rm E}
 + \sin \eta
|\,a_{3}\,\rangle_{\rm E} \nonumber \\
\hat\xi |\,a_{2}\,\rangle_{\rm E} = \cos \eta
|\,a_{2}\,\rangle_{\rm E} - \sin \eta |\,a_{4}\,\rangle_{\rm E} 
\end{eqnarray}
where $|\,a_{3}\,\rangle_{\rm E}$ and $|\,a_{4}\,\rangle_{\rm E}$ are 
the states orthogonal to $H_0$ and  
${}_{\rm E}\langle a_{3}\,|\,a_{4}\,\rangle_{\rm E} = 0$, yields
the same dependence of $Q$ and $B$ on $\eta$ as Type 2.
When $|\cos\eta|\neq 1$, this $\hat\xi$ does not satisfy Eq.~(\ref{koyuu}), 
which means that $|Q^{(2)}(\hat{\eta})|$ cannot be the minimum
of $|Q(\hat\xi)|$.
When $|\cos\eta|= 1$, $|Q^{(2)}(\hat{\eta})|$ is unity from Eq.(\ref{Qdayo}). 
Therefore, we can neglect $|Q^{(2)}|$ in determining $Q_{\rm min}(B)$.

Next, we consider the case when 
 the rank of $\nu\hat A - \lambda \hat B$
is $1$. Solving $\det(\nu\hat A - \lambda \hat B)=0$ in $H_{0}$,
we find 
that this case happens when $\lambda=0$ [since det$\hat{A} =0$ 
from Eqs.~(\ref{hatA})-(\ref{Qc})], or
\begin{equation}
 \frac{\lambda}{\nu}= \frac{A_{11} B_{22} + A_{22} B_{11} - 2 A_{12}
B_{12}}{B_{11} B_{22} - B_{12}^{2}}\equiv \kappa.
\end{equation}
The support of $\nu\hat A - \lambda \hat B$ is one-dimensional and 
let us write the corresponding pure state as $|a\rangle_{\rm E}\in H_0$.
 Let $|a^\perp \rangle_{\rm E} \in H_0$ be the 
state orthogonal to $|a\rangle_{\rm E}$. Using the same way as the
derivation of Eqs.~(\ref{type1}) and (\ref{type2}), 
Eq.~(\ref{koyuu}) implies that the matrix form of $\hat \xi$ in $H_0$ 
with respect to the basis $\{|a\rangle_{\rm E}, |a^\perp \rangle_{\rm E}\}$
should generally be written as
\begin{equation}
\hat \xi=\pm \pmatrix{1 & 0 \cr 0 & \cos\eta}\,.
\end{equation}
When $\lambda=0$, $|Q(\hat \xi)|$
is always unity (from Eq.(\ref{Qdayo})), and
is not relevant for the present 
problem of finding the minimum of $|Q(\hat \xi)|$.
The case $\lambda=\kappa\nu$ is relevant, and 
we call it Type 3. 
With the help of Eq.(\ref{Qdayo}), $Q(\hat{\xi})$ and $B(\hat{\xi})$ for 
Type 3 are 
given by functions of parameter $\eta$ as
\begin{eqnarray}
 Q^{(3,\pm)}(\eta) &=& \pm[_{\rm E}\langle a |\hat A|a \rangle_{\rm E} + \cos \eta
 _{\rm E}\langle
\, a^{\perp}| \hat A  |a^{\perp}  \rangle_{\rm E} ]
\nonumber \\
&=& \pm[{\rm Tr}(\hat A\hat{P}_a)(1-\cos\eta)+\cos\eta]
 \label{t31}
\end{eqnarray}
and from Eq.(\ref{Bda})
\begin{eqnarray}
  B^{(3,\pm)}(\eta) &=&\pm[ _{\rm E}\langle a |\hat B|a \rangle_{\rm E} + \cos \eta
 _{\rm E}\langle
\, a^{\perp}| \hat B  |a^{\perp}  \rangle_{\rm E}] 
\nonumber \\
&=& \pm[{\rm Tr}(\hat B\hat{P}_a)(1-\cos\eta)+\cos\eta{\rm Tr}(\hat
B)]
 \label{t32}
\end{eqnarray}
respectively, where 
\begin{equation}
\hat{P}_a\equiv |a \rangle_{\rm E} {}_{\rm E}\langle a |=
\frac{\hat{A}-\kappa\hat{B}}{{\rm Tr}(\hat{A}-\kappa\hat{B})}.
\end{equation}
$|Q^{(3,\pm)}|$ is plotted in Fig.~\ref{BQ} by dotted line.
In the figure, only $|Q^{(3,-)}|$ is plotted since $|Q^{(3,+)}|$
does not exist.

Finally, we consider the case where $\epsilon=0$. In this case,
 $Q_{{\rm min}}(B)$ can be directly found as
 \begin{equation}
 Q(B)=\frac{B}{\cos (\alpha + \theta)} \, ,
 \end{equation}
and $B_{0}=0$.

\begin{figure}
 \begin{center}
 \includegraphics[scale=0.7]{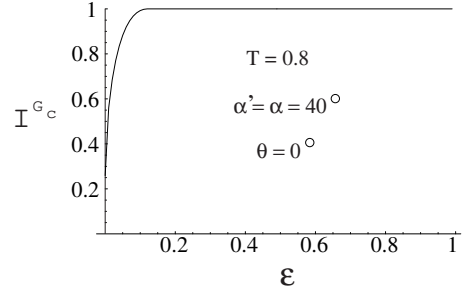}
 \caption{Eve's maximum information gain ${\rm I}^{\rm{Gc}}$ vs $\epsilon$ when $\alpha'= \alpha = 40^{\circ}$, $\theta = 0^{\circ}$, and $T=0.8$. \label{ig45}}
 \end{center}
\end{figure}

\begin{figure}
 \begin{center}
 \includegraphics[scale=0.7]{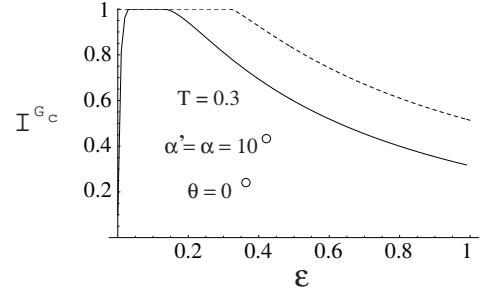}
 \caption{Eve's maximum information gain ${\rm I}^{\rm{Gc}}$ vs $\epsilon$ when $\alpha'= \alpha = 10^{\circ}$, $\theta = 0^{\circ}$, and $T=0.3$. The dashed
 line is the information bound based on Eq.~(\ref{S-bound}). \label{ig10}}
 \end{center}
\end{figure}

\begin{figure}
 \begin{center}
  \includegraphics[scale=1.0]{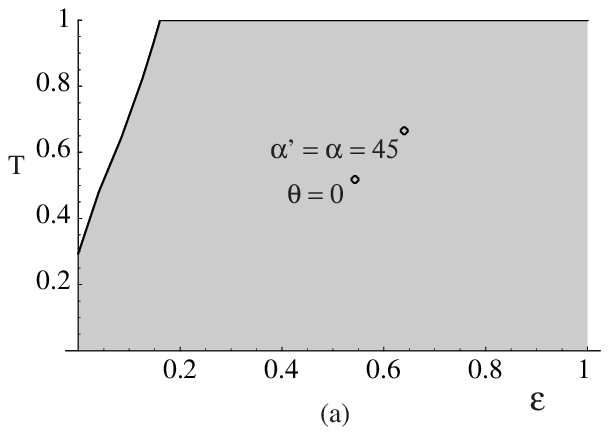}
  \includegraphics[scale=1.0]{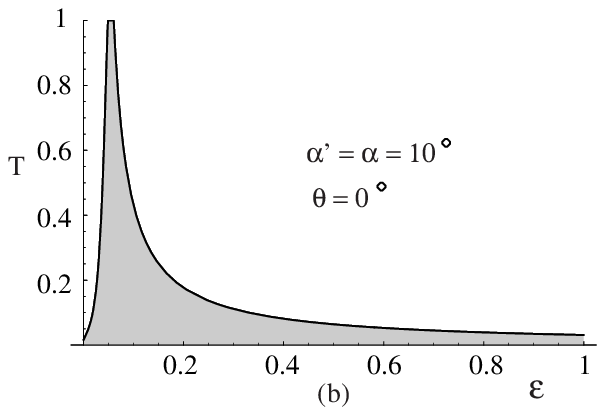}
 \caption{Eve's information gain is unity in the gray region
when (a) $\alpha'= \alpha = 40^{\circ}$, and $\theta =0$, and (b)
$\alpha'= \alpha = 10^{\circ}$ and $\theta =0$.
  \label{ig1}}
 \end{center}
\end{figure}

Now, since we have listed up all candidates for Eve's optimum operation,
we can obtain Eve's maximum information gain by picking up the optimum
one numerically.
We show some examples of Eve's information gain as a function of
$\alpha'$, $\alpha$, $T$, $\theta$, and $\epsilon$ in Fig.~\ref{ig45} and
Fig.~\ref{ig10}.
In these figures, we plot Eve's maximum information
gain as a function of $\epsilon$. Since the parameter $\epsilon$ 
corresponds to the noise or the error rate, it might be expected that
Eve's information gain increases when $\epsilon$ gets larger.
In Fig.~\ref{ig10}, however, Eve's information gain starts to 
decrease from the unity when the noise parameter $\epsilon$
exceeds a value around $0.13$.
In Fig.~\ref{ig1}, the shaded region represents the region where Eve's
information gain is unity when (a) $\alpha'= \alpha = 10^{\circ}$, and 
$\theta = 0^{\circ}$, and (b) $\alpha'= \alpha = 45^{\circ}$ and 
$\theta = 0^{\circ}$. We can see again in Fig.~\ref{ig1} (b) the
counter-intuitive behavior that the Eve's information gain
is lower in the region with larger $\epsilon$.

In order to explain the counter-intuitive behavior in Fig.~\ref{ig10},
let us consider the mutual information among the classical variables 
in Alice, Bob and Eve's sites. Let $A$, $B$, and $E$ be the random
variables describing Alice's bit value (0 or 1), Bob's measurement 
result ($\bar{0}$, $\bar{1}$, or inconclusive), and Eve's classical data
obtained from her attack, respectively. Let us consider the conditional 
mutual information 
${\rm{I(E;A,B|{\rm{c}})}} \equiv H(A,B|{\rm{c}})-
H(A,B|E,{\rm{c}})$, which is 
the mutual information between Eve and the joint system
of Alice and Bob for conclusive bits. Here $c$
represents the condition that $B=\bar{0}$ or $\bar{1}$,
 meaning that Bob's measurement result is conclusive.
The function $H(A,B|{\rm{c}})$
represents the entropy of the joint system
on condition that Bob obtains a conclusive result,
 and $H(A,B|E,{\rm{c}})$ stands for the conditional
entropy of the joint system averaged over Eve's variable $E$, 
namely, $H(A,B|E,{\rm{c}})\equiv \sum_y {\rm
Prob}(E=y)H(A,B|E=y,{\rm{c}})$.
Using the basic properties of the entropy function,
we can easily see that the following relation holds:
\begin{equation}
{\rm{I(E;A,B|{\rm{c}})}}
+{\rm{I(A;B|{\rm{c}})}}
= {\rm{I(A;B,E|{\rm{c}})}} + {\rm{I(B;E|{\rm{c}})}}.
\label{info-equality}
\end{equation}
The information ${\rm{I(E;A,B|{\rm{c}})}}$ is, hence, 
bounded as follows:
\begin{equation}
{\rm{I(E;A,B|{\rm{c}})}}\le {\rm{I(A;B,E|{\rm{c}})}} + {\rm{I(B;E|{\rm{c}})}}. 
\label{info-bound}
\end{equation}

The term ${\rm{I(A;B,E|{\rm{c}})}}$ in the right-hand side
means how well the joint
system of Bob and Eve can distinguish Alice's states on condition that Bob
obtains a conclusive result, and this term can be bounded from the 
fact that Alice's states are nonorthogonal, as follows.
With the help of the inequality
\begin{equation}
{\rm{I(A;B,E)}}\ge P_{\rm{conc}} {\rm{I(A;B,E|{\rm{c}})}}+
(1-P_{\rm{conc}}) {\rm{I(A;B,E|{\rm{inc}})}}\, ,
\end{equation}
where ``${\rm inc}$'' means $B={\rm inconclusive}$ and $P_{\rm{conc}}$
is the probability that Bob obtains the conclusive results. 
For $\theta =0$ and $\alpha ' =\alpha$, $P_{\rm{conc}}$ is given by
\begin{equation}
P_{\rm{conc}}= \frac{T}{4}\left[2 - (1 - \epsilon)
( \cos 2 \alpha+1)
\right] \, ,
\label{pconc1}
\end{equation}
we can bound ${\rm{I(A;B,E|{\rm{c}})}}$ as
\begin{equation}
{\rm{I(A;B,E|{\rm{c}})}}\le \frac{{\rm{I(A;B,E)}}}{P_{\rm{conc}}}
\le\frac{ 1-h\left(\frac{1-\sqrt{1-\cos^{2}\alpha}}{2}\right)}
{P_{\rm{conc}}}\, .
\label{A-BE-bound}
\end{equation}
The second inequality in Eq.~(\ref{A-BE-bound}) comes
from the optimum measurement on two nonorthognal pure states,
which was mentioned in Sec.~\ref{secs:protocol}.

The second term ${\rm{I(B;E|{\rm{c}})}}$
in the right-hand side of Eq.(\ref{info-bound})
 means how well Eve can control Bob's measurement outcomes,
$B=\bar{0}$ and $B=\bar{1}$. 
Since Bob's POVM elements 
$F_{\overline 0}$ and $F_{\overline 1}$ are nonorthogonal, Eve cannot 
control Bob's outcome as she please, and ${\rm{I(B;E|{\rm{c}})}}$
generally decreases as $\alpha$ decreases. 
As we discuss the detail in
Appendix \ref{upper}, ${\rm{I(B;E|{\rm{c}})}}$ is bounded by 
\begin{eqnarray}
& & {\rm{I(B;E|{\rm{c}})}}\le 1 \nonumber \\
&-&  h\left(\frac{1}{2}-
\frac{\sin\alpha}{4 (P_{\rm{conc}}/T)}
\sqrt{1-\left(\frac{2 (P_{\rm{conc}}/T)-1}{\cos\alpha}\right)^2} \right) \, .
\label{BE-bound}
\end{eqnarray}

Combining Eqs.~(\ref{info-bound}), (\ref{A-BE-bound}), and 
(\ref{BE-bound}), the mutual information ${\rm{I(E;A,B|{\rm{c}})}}$
is  bounded by
\begin{eqnarray}
& &{\rm{I(E;A,B|{\rm{c}})}} \le \frac{ 1- h\left(\frac{1-
\sqrt{1-\cos^{2}\alpha}}{2}\right)}{P_{\rm{conc}}} \nonumber \\
&+&1- h\left(\frac{1}{2}-\frac{\sin\alpha}{4 (P_{\rm{conc}}/T)}
\sqrt{1-\left(\frac{2 (P_{\rm{conc}}/T)-1}{\cos\alpha}\right)^2} \right)
 \nonumber \\
&\equiv& \nu 
\end{eqnarray}

After the transmission of $n$ pulses from Alice to Bob, 
we expect $n_{\rm conc}\equiv nP_{\rm conc}$ conclusive events on average, 
and Eve's information 
about Alice and Bob's bits for these events is
bounded by $nP_{\rm conc}\nu$. This quantity approaches
zero when $\alpha$ goes to zero. On the other hand, 
Alice and Bob obtain $n_{\rm cor}\equiv n P_{\rm{conc}}(1-e)$ correct bits,
where $e$ is the bit
error rate given by 
\begin{equation}
e = \frac{\epsilon}{2 - (1 - \epsilon) \, 
\left( \cos 2 \alpha + 1 \right)} \, ,
\label{s-error}
\end{equation}
for $\theta =0$ and $\alpha ' = \alpha$.
The number of correct bits $n_{\rm cor}$
does not necessarily approach zero when $\alpha$ goes to zero.
The information gain per one correct bit,  $\rm{I}^{Gc}_{S}$, 
cannot exceed $n_{\rm conc}\nu / n_{\rm cor}$, namely,
\begin{equation}
{\rm{I}^{Gc}_{S}} \le \frac{\nu}{1-e}\equiv
\rm{I}^{upper}_{S}\, . 
\label{S-bound}
\end{equation}
Since Shannon information gain ${\rm I}^{{\rm Gc}}_{{\rm S}}$
[Eq.~(\ref{IS})] and the information gain
${\rm{I}^{Gc}}$ [Eq.~(\ref{IR})] are connected by only one parameter $Q$, 
we can bound ${\rm{I}^{Gc}}$
through
$\rm{I}^{upper}_{S}$. In Fig.~\ref{ig10}, we plot this information bound
by the dashed line. The dashed line decreases as $\epsilon$ gets larger.

To summarize, the counter-intuitive behavior can be explained as follows.
When $\alpha$ is small, Alice's bit value cannot be guessed well 
from the outside since she encodes it into nonorthogonal states.
Bob's measurement results are also hard to guess since they 
come from nonorthogonal measurements. As a result, the mutual information 
between Alice and Bob, $\rm{I(A;B|c)}$, and Eve's information 
$\rm{I(E;A,B|c)}$, are both upper-bounded by the quantity $\nu$, which 
approaches zero when $\alpha$ goes to zero. On the other hand, 
when $\epsilon$ is large, the number of the conclusive bits is 
not small even when $\alpha$ is close to zero.
This means that Alice's bits and Bob's bits for the conclusive bits
are almost uncorrelated when $\alpha$ is close to zero.
Then, Alice and Bob
construct the correct bits by picking up 
the bits whose values accidentally coincide,
 through the encrypted communication over the classical channel.
In this case, the correct bits are essentially generated in this 
encrypted transmission, and Eve's information gain about them 
is very low. It is, however, not practical to perform
quantum key distribution in this region, 
since Alice and Bob must use many bits of the initially
shared secret key to determine the correct bits.

\begin{figure}
 \begin{center}
  \includegraphics[scale=1.0]{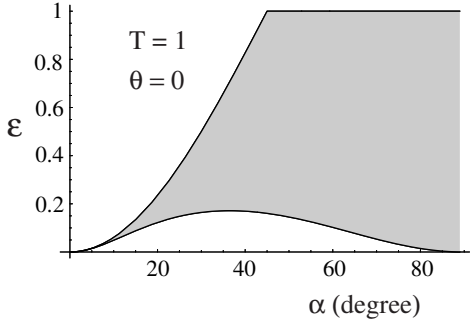}
 \caption{Eve's information gain is unity in the gray region $\{\alpha,
 \epsilon\}$ for $T=1$ and $\theta=0$.
  \label{ig1-2}}
 \end{center}
\end{figure}

Figure \ref{ig1-2} shows in gray the region where Eve's information 
gain is unity in
the  parameter region $\{\alpha, \epsilon\}$ for $T=1, 
\theta=0, \alpha'=\alpha$. The upper-left white region 
(small $\alpha$ and large $\epsilon$)
corresponds to the counter-intuitive behavior
 we have just discussed.
The figure shows another interesting behavior for small
$\epsilon$; Eve's
information gain is unity only when 
$\alpha$ is small or large,  and is not unity for 
 intermediate values of $\alpha$.
 For
$T<1$, the region for 
the full information gain for Eve has a similar 
shape to Fig.~\ref{ig1-2}, but is wider because the 
transmission loss gives advantage to Eve. In order to understand this 
behavior at
$T=1$, we consider the  following specific individual attack
that gives Eve the perfect information about the correct bits.

If Eve is sure that the state $|1\rangle(=|\sigma_\alpha\rangle)$
released by Alice always evolves to $|0\rangle(=|\sigma_{-\alpha}\rangle)$
and then reaches Bob, Eve can be sure that the value of the conclusive bit
surviving the error discarding procedure is 0. The simplest way to 
perform this attack is to rotate the system counter-clockwise by angle 
$2\alpha$ in the Bloch sphere (see Fig.~\ref{fig1}). To keep the symmetry, for 
half of the cases Eve rotates the system by angle $-2\alpha$. For the 
latter cases 
she can be sure that the value of the correct bit is $1$. 
In this attack, the averaged state $\rho^{\rm s}_1$ received by 
Bob when Alice has emitted $|1\rangle$ is given by 
$\rho^{\rm s}_1=(|\sigma_{-\alpha}\rangle\langle\sigma_{-\alpha}|
+|\sigma_{2\alpha}\rangle\langle\sigma_{2\alpha}|)/2=
\cos^2\alpha|\sigma_{\alpha}\rangle\langle\sigma_{\alpha}|
+\sin^2\alpha|\overline{\sigma_{\alpha}}\rangle\langle\overline{\sigma_{\alpha}}|$.
This attack is thus successful when $\epsilon=2\sin^2\alpha$, 
which coincides with the upper boundary in Fig.~\ref{ig1-2}.

The above ``rotating'' strategy can be improved by 
performing a weak measurement before the rotation, so that it can be used for 
smaller values of $\epsilon$. 
The weak measurement with outcomes $\{+,-\}$ 
 is described by the POVM $\{\hat A_{+},\hat A_{-}\}$, where
\begin{equation}
\hat A_{+}= (1-q) |x+\rangle\langle x+| + q|x-\rangle\langle x-|
\label{weak1}
\end{equation}
and
\begin{equation}
\hat A_{-}= q|x+\rangle\langle x+| + (1-q)|x-\rangle\langle x-|
 \,.
\label{weak2}
\end{equation}
The parameter $q$ $(0\le q \le 1/2)$
represents the weakness of the measurement.
If Alice has emitted $|j\rangle$ $(j=0,1)$, the outcome ``$+$" and ``$-$"
occur with probability 
$p_{j+}\equiv \langle j|\, \hat A_{+}|j\rangle$
and 
$P_{j-}\equiv \langle j|\, \hat A_{-}|j\rangle$,
respectively.
When this 
weak (projection) measurement produced outcome ``$+$", 
the angle $\beta$ (in the Bloch sphere) between the 
postmeasurement states 
$\sqrt{\hat A_{+}}|0\rangle$ and $\sqrt{\hat A_{+}}|1\rangle$
 is given by
\begin{equation}
\cos^2(\beta/2)=|\langle 0|\hat A_{+}|1\rangle|^2/ (p_{0+}p_{1+}).
\end{equation}
The outcome ``$-$" gives the same angle by the symmetry.
The angle $\beta$ is monotone increasing from $0$ to 
$2\alpha$ as a function of $q\in [0,1/2]$.

After this measurement, Eve rotates the system 
depending on the outcome. For ``$+$", she rotates it so
that $\sqrt{\hat A_{+}}|0\rangle$ becomes $|1\rangle$.
The state $\sqrt{\hat A_{+}}|1\rangle$ moves to 
$|\sigma_{\alpha+\beta}\rangle$ in this rotation.
When the outcome is  ``$-$",  she rotates it so
that $\sqrt{\hat A_{+}}|1\rangle$ becomes $|0\rangle$.
In this attack, the averaged state $\rho^{\rm s}_1$ received by 
Bob when Alice has emitted $|1\rangle$ is given by 
$\rho^{\rm s}_1=p_{1-}|\sigma_{-\alpha}\rangle\langle\sigma_{-\alpha}|
+p_{1+}|\sigma_{\alpha+\beta}\rangle\langle\sigma_{\alpha+\beta}|$.
This state can be written in the form 
$(1-\epsilon/2)|\sigma_{\alpha}\rangle\langle\sigma_{\alpha}|
+(\epsilon/2)|\overline{\sigma_{\alpha}}\rangle\langle\overline{\sigma_{\alpha}}|$
when 
\begin{equation}
\sin{\beta}/\sin{2\alpha}=p_{1-}/p_{1+}
\label{sspp}
\end{equation}
holds. In such a case, the value of $\epsilon$ is given by
\begin{equation}
\epsilon=1-\frac{\sin(2\alpha+\beta)}{\sin 2\alpha+\sin\beta}.
\label{eab}
\end{equation}
The equation (\ref{sspp}) has two solutions in $q\in [0,1/2]$,
$q=q_0(\alpha)$ and $q=1/2$. The case of $q=1/2$
corresponds to the simple rotating strategy described 
before. The strategy corresponding to the other solution
$q=q_0(\alpha)$ turns out to give the lower boundary in
Fig.~\ref{ig1-2}. The entire gray region in Fig.~\ref{ig1-2}
is then covered by simply mixing the two strategies
corresponding to $q=q_0(\alpha)$ and $q=1/2$.

The behavior for small $\epsilon$ shown in Fig.~\ref{ig1-2}
will now be understood as follows. When $\alpha$ is small,
the simple rotating strategy with the
 small rotation angle $2\alpha$ causes only a small 
disturbance, and it can be used to obtain small $\epsilon$.
When $\alpha$ is large and the two states can be well distinguished,
 the improvement of the strategy by the measurement
 is quite effective to reduce $\epsilon$. For intermediate 
values of $\alpha$, $p_{0+}/p_{1+}$ cannot be close to 
zero. Since $p_{0+}=p_{1-}$, the angle $\beta$ cannot be
close to zero either, due to the constraint (\ref{sspp}).
Consequently, as is implied by Eq.~(\ref{eab}), 
even the improved strategy cannot reduce $\epsilon$ close to
zero.

\section{Secret Key Gain} \label{key-gain}

In this section, we show the calculation of secret key gain and the 
optimum angle of Alice's states to obtain optimum secret key gain.
For simplicity, we assume that $\alpha=\alpha '$ and $\theta=0$.
The calculation for more general cases is straghtforward.

We define the secret key gain $G$ as the net growth of
the secret key per one pulse. Alice and Bob can increase the 
length of the secure secret key in the region where $G$ is positive.

In the calculation of key gain, we assume the error correction protocol
 whose number of
redundancy bits used to detect the errors of the raw key are 
equal to the Shannon limit, i.e.,
$n h(e)$, where $n$ is the length of the raw key,
 $e$ is the bit error rate of the raw key given by Eq.~(\ref{s-error}), and 
$h(e)$ is entropy function. 
We further assume that the privacy amplification \cite{privacy amp}
is applied to the correct bits and to the flipped bits independently.
Since Eve
may not be allowed to adopt the strategy that is simultaneously 
optimum both for correct bits and
flipped bits, this protocol may be overkill but never
 underestimates Eve's ability.
 
For the privacy amplification of the correct bits, 
we use Eve's information estimated via the method described
in the previous sections.
From these bits Alice and Bob  
obtain $G^{\rm new}_{c}$ bits per pulse of new secret key,
which is written as
 \begin{equation}
G^{\rm new}_{c} \equiv P_{{\rm conc}} (1-e) (1-{\rm I}^{{\rm Gc}}) - 
(s_{c}  /n_{{\rm total}})\, \label{gaintempc}
\end{equation}
where $P_{{\rm conc}}$ is given by Eq.~(\ref{pconc1}), $n_{{\rm total}}$ is
the number of pulses emitted by Alice, and 
$s_{c}$ is the security parameter for correct bits. Eve's information 
about the secret key obtained from the correct bits
is less than $2^{-s_{c}} / \ln 2$ (bits).

For flipped bits, Alice and Bob 
obtain $G^{\rm new}_{f}$ bits per pulse of new secret key,
which is written as
 \begin{equation}
G^{\rm new}_{f} \equiv P_{{\rm conc}} e (1-{\rm I}^{{\rm Gf}}) - (s_{f} /n_{{\rm total}})\, .
\label{gaintempe}
\end{equation}
Eve has information less than $2^{-s_{f}} / \ln 2$ (bits) about the secret key from the flipped bits.
The information gain for flipped bits $\rm{I}^{{\rm Gf}}$
can directly be obtained by replacing $\theta$ with 
$- 2\alpha-\theta$ in the formula in Sec.~\ref{sec:sym}. 

Noting that the redundancy bits used in the error
correction protocol are encrypted by 
consuming $P_{{\rm conc}}h(e)$ bits per pulse of the 
initially shared
secret key, we obtain
the expression of the secret key gain $G$ as follows,
 \begin{eqnarray}
G &\equiv& G^{\rm new}_{c} + G^{\rm new}_{f} - P_{{\rm conc}} h(e)
\nonumber \\
&=& P_{{\rm conc}} \left[(1 - e )(1-{\rm I}^{{\rm Gc}})  
+e(1-{\rm I}^{{\rm Gf}}) - h(e) \right] \nonumber \\
 &-&  (s_{c}
+s_{f})  /n_{{\rm total}}\, .
\label{gaintemp}
\end{eqnarray}
 In the limit of long key, i.e.,
  $n_{{\rm total}} \to \infty$, $G$ is reduced to 

 \begin{equation}
G = P_{{\rm conc}} \left[(1 - e )\,(1-{\rm I}^{{\rm Gc}})   + 
e (1-{\rm I}^{{\rm Gf}})   - h(e) \right] \, .
\label{gain}
\end{equation}

\begin{figure}
 \begin{center}
 \includegraphics[scale=1.3]{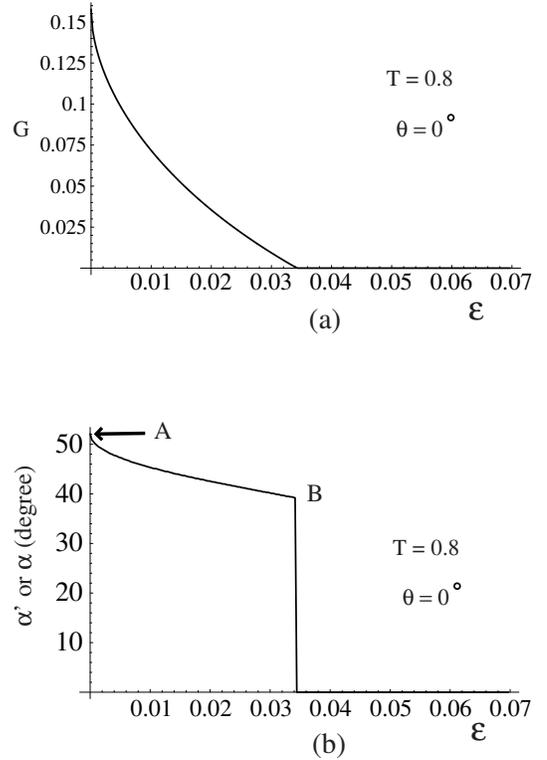}
\caption{(a) The optimized secret key gain per pulse and (b) optimal angle $\alpha^{\circ}$ as a function of $\epsilon$. $T = 0.8$, $\alpha' = \alpha $ and $\theta = 0^{\circ}$. ``A'' represents the angle when $\epsilon = 0$ and ``B'' represents the angle when the key gain vanishes.  \label{opt8}}
  \end{center}
\end{figure}

In B92 protocol, if $\alpha' (= \alpha)$ is changed, the secret
key gain will be changed. So, Alice and Bob should optimize this
angle to obtain higher secret key gain. In Fig.~\ref{opt8}, we optimize
$\alpha' \,(= \alpha)$ for fixed values of $T$ and $\epsilon$
\cite{gene} to obtain high secret key gain, and plot the key gain and the
optimized angle as a function of $\epsilon$.
 The points ``A'' and ``B''
represent the cases when $\epsilon = 0$ and when the key gain 
 vanishes, respectively. In the figure, as $\epsilon$ increases,
the optimum angle tends to be smaller, and the region where optimum
angle $\alpha' =0$ implies that our protocol does not work. In the
figure, there is only a negligible contribution of flipped bits to the key gain.

\begin{figure}
 \begin{center}
 \includegraphics[scale=0.6]{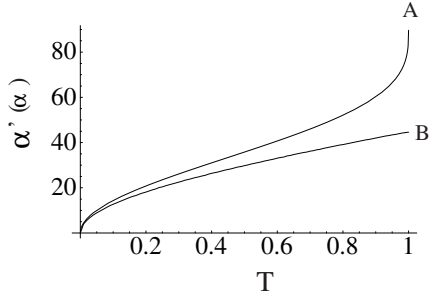}
  \caption{The $T$ dependence of``A'' and ``B''. \label{ab}}
 \end{center}
\end{figure}

\begin{figure}
 \begin{center}
 \includegraphics[scale=0.6]{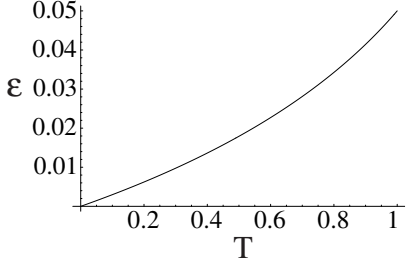}
  \caption{The maximum value of $\epsilon$ up to which the secret key gain is positive as a function of $T$. \label{maxi}}
 \end{center}
\end{figure}

To investigate the $T$ dependence on the optimum angle which are 
represented as ``A'' and ``B'', we plot these angles as a function of
$T$ in Fig.~\ref{ab}. For ``A'', the
key gain has a simple expression
\begin{equation}
G = \frac{T}{4} (1 - \cos 2 \alpha') \left[ 1 - \log_{2} \left(2 - \left( \frac{\cos \alpha' - 1 + T}{T \cos \alpha'} \right)^{2} \right) \right] \, . 
\end{equation}
In Fig.~\ref{ab}, it is seen that as $T$ gets larger,
the optimum angle tends to get larger. 
In Fig.~\ref{maxi}, 
we plot $\epsilon$ of ``B''
as a function of $T$. 
The region below the curve means the parameter region of $\{T,\epsilon \}$
where a secure key can be produced in our protocol.

\begin{figure}
 \begin{center}
 \includegraphics[scale=0.8]{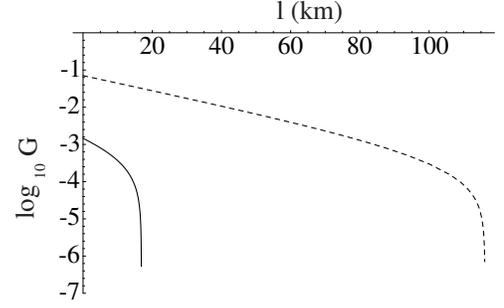}
 \caption{Comparison of B92 (solid line) with BB84 (dotted line) in the case where $L_{c} =0.2$(dB), $L_{r} = 1$(dB), $\nu = 2 \times 10^{-4}$(per pulse), and $\eta = 18$ (\%). The dotted line represents BB84 and solid line represents B92 protocol. \label{vs}}
 \end{center}
\end{figure}

 Finally, to compare the security of B92 with that of BB84 assuming an ideal
single-photon source, we consider the dependence of $G$ as a
function of distance between Alice and Bob, 
while imperfection factors such as dark 
counting of detectors and 
losses of optical fibers are fixed. For the purpose of this comparison,
we consider the case where Bob uses photon counters which 
can discriminate no photon, single-photon, and higher number of 
photons. Let $l$, $L_{c}$, $L_{r}$, $\nu$, 
and $\eta$ be the length of transmission line (km), the channel loss (dB/km), 
the receiver loss (dB), the mean dark count per pulse (which is given by $\nu = \tau_{{\rm res}} R_{{\rm dark}}$ where $R_{{\rm dark}}$ is the dark counting rate and $\tau_{{\rm res}}$ is the resolution time of the detector and 
 electronic circuitry), and the detection efficiency, respectively. 
 In this case, $T$ is given by \cite{barnett}
\begin{equation}
T = {\rm e}^{- \nu} \, \eta \, 10^{- (l L_{c} + L_{r})/10} + 
{\rm e}^{- \nu}\,\nu (1 - 10^{- (l L_{c} + L_{r})/10}) \, ,
\end{equation}
where we assume that the dark counts are Poissonian events.
From this equation and 
 \begin{eqnarray}
 \tilde \rho_{i}^{{\rm s}} &=& \frac{{\rm e}^{- \nu} \, \eta \,
10^{- (l L_{c} + L_{r})/10}}{T} |i\rangle\langle i| \nonumber \\
&+& 
\frac{{\rm e}^{- \nu}\,\nu (1 - 10^{- (l L_{c} + L_{r})/10})}{T}
\frac{\textbf 1}{2}
\nonumber \\
&=& (1-\epsilon) |i\rangle\langle i|+\frac{\epsilon}{2}
 \, {\textbf 1} \,,
 \end{eqnarray}
$\epsilon$ can be written as 
\begin{equation}
\epsilon = \frac{\,{\rm e}^{- \nu}\,\nu (1 - 10^{- (l L_{c} 
+ L_{r})/10})}{T} \, .
\end{equation}
In Fig.~\ref{vs}, we plot $\log_{10} G$ as a function of distance $l$(km). 
In the figure, we put $\alpha (\alpha') = 11$(degree) which is almost optimum, and    
 experimental data is taken from KTH \cite{KTH}. For the secret key gain of BB84 protocol
 in the figures, we used the formula for the single-photon case in \cite{norbert2}, 
 \begin{eqnarray}
 G &=& \frac{T}{2} [1 - \log_{2} (1+4e-4e^{2})+e\log_{2} e \nonumber \\
 &+& (1-e)\log_{2} (1-e)] \, ,
 \end{eqnarray}
where $e=\nu/(2T)$,
assuming that the errors stem from the dark counting.
In Fig.~\ref{vs}, it is seen that B92 protocol is far less efficient, 
which is mainly because a low transmission rate $T$ directly gives an 
advantage to Eve in the B92 scheme using photon polarization.

\section{Summary and Discussion}\label{summary} 

\begin{figure}
 \begin{center}
 \includegraphics[scale=0.7]{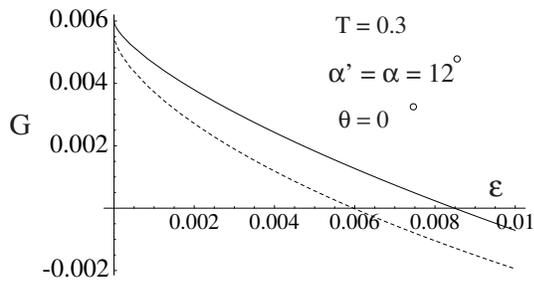}
 \caption{The secret key gain vs $\epsilon$ in the case of $T = 0.3$, $\alpha' = \alpha = 12^{\circ}$ and $\theta = 0^{\circ}$. Eve's information is estimated by Shannon entropy (solid line) and ${\rm I}^{{\rm Gc}}$ (dotted line). The estimation by Shannon entropy is more efficient. \label{shannon}}
  \end{center}
\end{figure}

Throughout this paper, we have estimated Eve's information gain 
${\rm I}^{{\rm Gc}}$. But as Bennett {\it et al.} pointed out in Sec.~VI 
of \cite{privacy amp}, if eavesdropping is independently done for each pulse, privacy amplification using information
gain based on the expected collision probability is overkill. In such case, 
it follows that it may be sufficient to estimate Eve's information gain by 
Shannon entropy. In the case where Eve employs 
the symmetric von Neumann measurement, the proof in \cite{privacy amp} assures 
that Shannon entropy is enough for the estimation. 
Maximizing Eve's information gain with respect to Shannon entropy is equivalent to the
problem of minimizing $|Q|^2$, which is the same problem as the maximization of Eve's
information gain ${\rm I}^{{\rm Gc}}$. So, our method described in the previous sections can
be directly applied to this problem. We plot an example of the secret key gain for some cases when we
estimate Eve's maximum information gain by Shannon entropy in Fig.~\ref{shannon}. We see that the key gain is higher than the case where Eve's information 
is estimated by ${\rm I}^{{\rm Gc}}$. 
 
Another candidate which may make the key gain higher is error reconciliation protocols. There may be the case where error discarding protocols \cite{error discarding} are more efficient than the error correction protocol, because 
the redundancy bits consumed in a discarding protocol may be less, for
Alice and Bob do not need to keep the discarded bits secret in the error discarding protocol. 
In the calculation of all figures in Sec.~\ref{key-gain}, we have
checked that there is no contribution of the flipped bits to the 
secret key, which implies
that the use of an error discarding protocol may be more efficient than that
of an error correction protocol. 

Since the density matrices which Alice emits are fixed as well as
those which are received by Bob, the problem of our estimation of 
quantum key distribution over a realistic channel is reduced to the 
problem that how much information Eve can extract on the 
condition that both initial density matrices $\rho_{i}$ ($i = 0, 1$) 
and final density matrices $\rho_{i}'$ are fixed.
In the realistic situation, however, since the number of pulses
$n_{\rm total}$ which
are emitted by Alice is finite, we cannot uniquely identify 
the final density matrices. In such a situation, by considering the 
standard deviation of Bob's data, we can pick up the candidates of 
the final density matrices, and we can estimate Eve's maximum information
gain by comparing information gain for each candidates. 

To summarize this paper, we derived a formula that estimates
Eve's maximum information when Alice and Bob perform B92 protocol
using two nonorthogonal single-photon polarization states.
We have assumed that Alice can emit a
single photon and Eve employs individual attack. 
The problem is equivalent to ask how much information 
can be extracted while simulating a noisy channel 
through which Alice communicates with Bob by sending
the two nonorthogonal states. The dependence of 
the maximum information on the noise parameter shows 
 nontrivial behavior. It decreases as the noise
increases in some parameter regions. For a small noise rate,
Eve can extract perfect information in the case where the angle of
Alice's two states is very small or very large, while she cannot
extract perfect information for intermediate angles.
Using the formula,
we plot the secret key gain as a function of various parameters reflecting
intervention by Eve. We also investigated the optimum 
angle between the two states to obtain large secret
key gain, and showed the region of the parameters where this key gain
is positive. 
A comparison of our protocol with BB84 protocol
 shows that B92 protocol with single photon
polarization is less efficient than BB84 protocol, which is mainly because
a low transmission rate $T$ directly gives an advantage to
 Eve in the B92 scheme using photon polarization.
This drawback may be compensated in the original B92 protocol
using two nonorthogonal coherent light, or in the 4+2 protocol
\cite{imoto} that uses the switching of the basis as in BB84 while 
retaining the nonorthgonality as a freely adjustable parameter.
 We leave the security problems of such protocols 
to future studies. 
 
We have assumed throughout this paper that eavesdropping is independent for 
each pulse. Even within the independent quantum operation, Eve can 
obtain a correlation among her eavesdropping outcomes by choosing her
strategy depending on the outcomes which have been already obtained.
The estimation against such attack is important for the realistic purpose, 
but we also leave this problem for future studies, along with the security
against collective attack or coherent
attack. 

\acknowledgements
This work was partly supported by a Grant-in-Aid for Encouragement of 
Young Scientists (Grant No. 12740243) and a Grant-in-Aid for Scientific 
Research (B) (Grant No. 12440111) by Japan Society of the Promotion of 
Science. 

\appendix
\section{Symmetrization of Eve's strategy}\label{apA} 

In this appendix, we construct a symmetrized strategy 
$U^{\rm s}$, which possesses higher symmetry and 
the same power as the actual strategy $U$,
namely,
$|Q(U^{\rm s})|=|Q(U)|$.
The problem of minimizing $|Q|$ over 
possible $U$ is then simplified to 
the minimization over possible $U^{\rm s}$.
Similar discussion is seen in
Appendix A of \cite{slu}.
In addition to the  symmetry, this reduction 
makes the analysis simpler because 
 the states delivered to Bob by the strategy $U^{\rm s}$
are completely specified by the observed quantities.

First, consider a strategy $U^\prime$ in which 
Eve conducts an additional (nondestructive) measurement of the photon number
 just after the original strategy of $U$ satisfying Eqs.~(\ref{constraint}) 
 and (\ref{density}).
 After this measurement, if the
outcome of this measurement is one photon, she sends the projected state to Bob, but if
the outcome is no photon or more than one photon, she always sends
the vacuum state $|\,{\rm vac}\,\rangle$. 
The states $\rho_{j}^\prime (j=0,1)$ delivered to Bob in this strategy are given by
\begin{equation}
\rho_{j}^\prime = P_1\rho_{j}P_1 + (1 - T_j)\,|\,{\rm vac}\,\rangle \langle{\rm vac}\,|,
\end{equation}
where $P_1$ is the projection onto the subspace $H_{1}$ of one photon and 
\begin{equation}
T_j\equiv 1-{\rm Tr}[F_{\rm V}\rho_j] \, ,
\end{equation}
which is the probability that Bob detects a single-photon.
Since ${\rm Tr}(F_\mu\rho_j^\prime)={\rm Tr}(F_\mu\rho_j)$ $(\mu=0,1,\bar{0},
\bar{1}; j=0,1)$, the strategy $U^\prime$ satisfies the condition 
Eq. (\ref{constraint}), and the additional measurement only gives Eve extra
information,
$|Q(U^\prime)|\le|Q(U)|$ holds. Hence, in finding the minimum of
$|Q(U)|$ under the
condition Eq.~(\ref{constraint}), we are allowed to assume the additional 
assumption that $\rho_j$ is equal to $\rho_{j}^\prime$ and is written as 
\begin{equation}
\rho_{j} = T_j \tilde{\rho}_{j} + (1 - T_j)\,|\,{\rm vac}\,\rangle \langle{\rm vac}\,|,
\end{equation}
where $\tilde{\rho}_{j} \equiv  P_1\rho_{j}P_1 /T_j$ is a normalized density operator in $H_{1}$.

Next, take a basis of the signal space $H$ that includes 
$| \, z + \rangle$, $| \, z - \rangle$, and $|\,{\rm vac}\,\rangle$
 as elements,
and take a basis of Eve's probe space that includes $|\,w\,\rangle$.
The product states of the two bases form a basis of the combined system, 
and let us denote it as ${\rm W}$. 
Let $Z_{{\rm W}}$ be the transformation which replaces all the vector
 components
and the elements of matrices in the basis ${\rm W}$ with their complex conjugates. 
Since the initial states $|\,j\,\rangle \,|\,w \rangle_{\rm E} (j=0,1)$ and Bob's measurement $\{F_\mu\}$ are invariant under the transformation $Z_{{\rm W}}$, and the probabilities of events in quantum mechanics are given by moduli squared of vector inner products, 
the strategy given by the unitary operator $Z_{{\rm W}} U Z_{{\rm
W}}^{-1}$ gives the same amount of the information gain to Eve
 as the strategy $U$.

Similarly, let $\sigma_z\equiv |z+\rangle\langle z+|-|z-\rangle
\langle z-|$
 be the $\pi$ rotation around the $z$ axis in the Bloch sphere, 
and $R$ be the extension of $\sigma_z$ to the entire space ($R \equiv
( \sigma_{z} \oplus F_{{\rm V}}) \otimes 1_{\rm E}$, where subscript E
 means Eve's probe space). 
Since the changes in the initial states $|\,j\,\rangle \,|\,w \,
 \rangle_{\rm E} (j=0,1)$ 
and Bob's measurement
$\{F_\mu\}$ under the transformation $R$
is just the interchange of the bit values $0$ and $1$,
the strategy given by the unitary operator $R U R^{-1}$ 
gives the same amount of the information gain to Eve
 as the strategy $U$.

 Since the four strategies $U$, $Z_{{\rm W}} U Z_{{\rm W}}^{-1}$, $R U
R^{-1}$, and
$R Z_{{\rm W}} U Z_{{\rm W}}^{-1} R^{-1}$
give the same amount of information gain, the strategy 
in which Eve randomly chooses one of the four unitary operations 
also gives the same amount of information gain. In other words,
if we write the unitary operator corresponding to this randomized
strategy as $U^{\rm s}$, then $|Q(U^{\rm s})|=|Q(U)|$ holds.
The states $\rho_{0}^{\rm s}$ and $\rho_{1}^{\rm s}$ delivered to Bob in this strategy are given by
 \begin{eqnarray}
 \rho_{0}^{\rm s} &=&  \frac{1}{4}({\rho}_{0} + Z_{{\rm W}} \rho_{0}
Z_{{\rm W}}^{-1} + R \rho_{1} R^{-1} \nonumber \\
 &+& Z_{{\rm W}} R \rho_{1} R^{-1} Z_{{\rm W}}^{-1}) \nonumber \\
 &=&T\tilde{\rho}_0^{\rm s}+(1-T) |{\rm vac}
 \rangle\langle {\rm vac}|
 \label{A1}
 \end{eqnarray}
 and
 \begin{eqnarray}
 \rho_{1}^{\rm s} &=&  \frac{1}{4}({\rho}_{1} + Z_{{\rm W}} \rho_{1}
Z_{{\rm W}}^{-1} + R \rho_{0} R^{-1} \nonumber \\
 &+& Z_{{\rm W}} R \rho_{0} R^{-1} Z_{{\rm W}}^{-1}) \nonumber \\
 &=&T\tilde{\rho}_1^{\rm s}+(1-T) |{\rm vac}
 \rangle\langle {\rm vac}|,
  \label{A2}
 \end{eqnarray}
 where
\begin{eqnarray}
\tilde{\rho}_0^{\rm s}&=&\frac{1}{4T}[T_0(\tilde{\rho}_{0} + Z_{{\rm W}} \tilde{\rho}_{0}
Z_{{\rm W}}^{-1}) + T_1(R \tilde{\rho}_{1} R^{-1} \nonumber \\
 &+& Z_{{\rm W}} R \tilde{\rho}_{1} R^{-1} Z_{{\rm W}}^{-1})],
  \label{A3}
 \end{eqnarray}
\begin{equation}
  \tilde{\rho}_{1}^{\rm s} = \sigma_z \tilde{\rho}_0^{\rm s} \sigma_z\,,
\label{A4}
\end{equation}
and
\begin{equation}
T\equiv\frac{T_0+T_1}{2}=1-{\rm Tr}[F_{\rm V}(\rho_0+\rho_1)/2]\,.
 \label{A5}
\end{equation}
From these expressions, we have
\begin{equation}
{\rm Tr}[\sigma_y\tilde{\rho_{j}}^{\rm s}]
=0,
\label{ycompo}
\end{equation}
where $\sigma_y\equiv i|z-\rangle\langle z+|-i|z+\rangle\langle z-|$.
Eqs.(\ref{A1}) and (\ref{ycompo}) implies that $\tilde{\rho}_{j}^{\rm s}$ are written in the Bloch sphere as in Fig.~\ref{fig4}. 
Taking the parameters $\theta$ and $\epsilon$ as in  Fig.~\ref{fig4}
(the sign of $\theta$ is positive in the clockwise),
$\rho_{j}^{\rm s}$ can be written in a diagonalized form as in Eqs.~(\ref{den1}) and (\ref{den2}).

\section{Optimization of Eve's apparatus}\label{apB} 
In this appendix, we derive Eq.~(\ref{koyuu}).
In order to determine the function $Q_{\rm min}(B)$, we use Lagrange's
method of undetermined multipliers. $B_0$ is the maximum of 
$B(\hat{\xi})$ under the condition that $Q(\hat{\xi})=0$.
Hence, if $B(\hat{\xi})=B_0$ and $Q(\hat{\xi})=0$, such $\hat{\xi}$
satisfies
\begin{equation}
\delta(B(\hat{\xi})-\nu Q(\hat{\xi})) =0
\label{lag0}
\end{equation}
for any variation of $\hat{\xi}$.
$Q_{\rm min}(B)$ for $B_0<B \le B_{\rm
max}$ is determined by  minimizing  $Q^2(\hat{\xi})$
under the condition that $B(\hat{\xi})=B$.
Suppose that  $Q^2(\hat{\xi})$ takes
its minimum at $\hat{\xi}$ under the condition that $B(\hat{\xi})$ is
fixed. Then, at least one of the following conditions holds, 
namely,
\begin{equation}
\delta B(\hat{\xi})=0
\label{lag1}
\end{equation}
 for any variation of $\hat{\xi}$, or,
there exists real $\lambda^\prime$ such that 
\begin{equation}
\delta
(Q^2(\hat{\xi}) -
\lambda^\prime B(\hat{\xi}))=0
\label{lag2}
\end{equation}
 for any variation of $\hat{\xi}$.
Since $Q(\hat{\xi})\neq 0$ in the considered region,
the condition (\ref{lag2}) is equivalent to 
\begin{equation}
\delta
(Q(\hat{\xi}) -
\lambda B(\hat{\xi}))=0 \, ,
\label{lag3}
\end{equation}
where $\lambda \equiv \lambda'/2Q$.
The three conditions (\ref{lag0}), (\ref{lag1}), and (\ref{lag3}) can be cast into 
a common form, namely,
 there exist real $\nu$ and  $\lambda$ 
[$(\nu,\lambda)\neq (0,0)$] such that 
\begin{equation}
\delta
(\nu Q(\hat{\xi}) -
\lambda B(\hat{\xi}))=0
\label{lagf}
\end{equation}
 for any variation of $\hat{\xi}$.
In determining $B_0$ and $Q_{\rm min}(B)$,
it is sufficient to consider only 
the operator $\hat{\xi}$ satisfying the above condition. 

Since $\hat{\xi}$ is a real orthogonal matrix in the basis ${\rm W}$, any
variation of $\hat{\xi}$ can be written as 
$\hat \xi \rightarrow \hat \xi \, (1 + \hat N \delta )$,
where $\hat N$ is an arbitrary real antisymmetric matrix 
in ${\rm W}$ and
$\delta$ is an infinitesimal real number.
Then, the condition (\ref{lagf}) means
 \begin{equation}
  {\rm Tr}\left[(\nu\hat A - \lambda \hat B) \, \hat \xi \, \hat N \right]
= 0  
 \end{equation}
for any  $\hat N$ whose matrix form in ${\rm W}$
is real and antisymmetric. This condition is satisfied
if and only if 
$(\nu\hat A - \lambda \hat B) \, \hat \xi$ is 
a real symmetric matrix in ${\rm W}$. Since 
$(\nu\hat A - \lambda \hat B)$ is 
a real symmetric matrix and 
$\hat \xi$ is a real orthogonal matrix in ${\rm W}$, we have
\begin{equation}
(\nu\hat A - \lambda \hat B) \, \hat \xi
=\hat{\xi}^{-1} (\nu\hat A - \lambda \hat B)
\end{equation}
and hence
\begin{equation}
[(\nu\hat A - \lambda \hat B)^2,\hat{\xi}]=0.
\end{equation}

\section{The upper bound of ${\rm{I(B;E|{\rm{c}})}}$}\label{upper} 
In this appendix, we derive an upper bound of ${\rm{I(B;E|{\rm{c}})}}$
in Eq.~(\ref{BE-bound}), assuming that Eve freely prepares quantum 
states and sends them to Bob. We only require that Eve should not change 
the values $T$ and $P_{\rm conc}$. Hence, Eve's strategy is generally 
described as she sends a single photon with its polarization 
in the state $\rho_\gamma$ with probability
$P(\gamma)T$, and she sends the vacuum with probability $1-T$.
The probability $P(\gamma)$ should satisfy 
\begin{equation}
\sum_{\gamma}P(\gamma)=1 \, ,
\label{normalize}
\end{equation}
and $\rho_\gamma$ and $P(\gamma)$ should satisfy 
\begin{equation}
\sum_{\gamma}P(\gamma) P({\rm{c}}|\gamma)=P_{\rm{conc}}/T \, .
\label{const2}
\end{equation}
Here $P({\rm{c}}|\gamma)$ is the probability of obtaining 
conclusive results for the state $\rho_\gamma$, which is 
given by 
\begin{eqnarray}
P({\rm{c}}|\gamma)&=&
{\rm{Tr}}\left[(F_{\overline{0}}+F_{\overline{1}})\rho_\gamma \right] 
\nonumber \\
&=&{\rm{Tr}}\left[\left(\cos^2\frac{\alpha}{2}|z+\rangle
\langle z+|+\sin^2\frac{\alpha}{2}|z-\rangle\langle z-|\right)\rho_\gamma
\right]  \nonumber \\
&=&\frac{1}{2}\left[1-\cos\alpha{\rm{Tr}}(\sigma_z \rho_\gamma) \right]\, .
\label{d-POVM1}
\end{eqnarray}
The mutual information ${\rm{I(B;E|{\rm{c}})}}$ can be written as
\begin{eqnarray}
{\rm{I(B;E|{\rm{c}})}}&=& H(B|{\rm{c}})- H(B|E,{\rm{c}}) \nonumber \\
&\le&1-\left(\frac{T}{P_{{\rm conc}}}\sum_{\gamma} P(\gamma) P({\rm{c}}|\gamma)
 H(B|\gamma ,{\rm{c}})\right) \, .
\label{H2}
\end{eqnarray}
The entropy $H(B|\gamma ,{\rm{c}})$ is given by
\begin{eqnarray}
H(B|\gamma
,{\rm{c}})
&=&h\left(\frac{1}{2}-\frac{\left|P(\overline0|\gamma,{\rm{c}})
-P(\overline1|\gamma,{\rm{c}})\right|}
{2}\right) \nonumber \\
&=&h\left(\frac{1}{2}-\frac{\left|
{\rm{Tr}}[(F_{\overline{0}}-F_{\overline{1}})\rho_\gamma]\right|}
{2P({\rm{c}}|\gamma)}\right).
\end{eqnarray}
Using 
\begin{equation}
F_{\overline{0}}-F_{\overline{1}}=\frac{\sin\alpha}{2}\sigma_{x} 
\label{d-POVM2}
\end{equation}
and the relation ${\rm{Tr}}(\sigma_z \rho_\gamma)^2+
{\rm{Tr}}(\sigma_x \rho_\gamma)^2\le 1$, we obtain
\begin{eqnarray}
H(B|\gamma ,{\rm{c}})&\ge& h\left(\frac{1}{2}-
\frac{\sin\alpha}{4P({\rm{c}}|\gamma)}
\sqrt{1-
\left(\frac{1-2P({\rm{c}}|\gamma)}{\cos\alpha}\right)^2} \right) 
\nonumber \\
&\equiv& g\left(P({\rm{c}}|\gamma)\right)
\, ,
\label{h-bound1}
\end{eqnarray}
Since the function $x\rightarrow xg(x)$ 
is convex (we have confirmed this
numerically), 
${\rm{I(B;E|{\rm{c}})}}$ in Eq.~(\ref{H2}) is maximum when 
$P({\rm{c}}|\gamma)=P_{\rm{conc}}/T$ for all $\gamma$,
 and
we obtain the result
\begin{equation}
{\rm{I(B;E|{\rm{c}})}}\le 1-g\left(P_{\rm{conc}}/T\right),
\end{equation}
which is Eq.~(\ref{BE-bound}).



\begin{thebibliography}{2}
\bibitem{note} Although the
Appendix B of \cite{slu} claims that the expected Renyi entropy is maximized 
by the symmetric von Neumann measurement, they have also proved in their proof 
an inequality, which reads $2\langle P_c \rangle -1\le
1-|Q|^2$ in our notation, showing that the expected collision probability 
$\langle P_c \rangle$ is also maximized by the symmetric von Neumann measurement.
\bibitem{gene} In general, there exist
correlations among $\alpha'$, $\alpha$, $\theta$, and $\epsilon$. 
But since we can calculate Eve's information gain for any $\alpha'$,
$\alpha$, $T$, $\theta$, and $\epsilon$, even if such a correlation
exists, Alice and Bob can maximize the secret key gain by taking this
correlation into consideration.
\end{thebibliography}
\end{document}